\documentstyle[12pt,psfig,a4]{article}
\begin{document}
\title{
Physics and Astrophysics of Strange Quark Matter
}

\author{
Jes Madsen \\
\em{Institute of Physics and Astronomy, University of Aarhus} \\
\em{DK-8000 {\AA}rhus C, Denmark}
}

\maketitle

\begin{abstract}
3-flavor quark matter (strange quark matter; SQM) can be
stable or meta\-stable for a wide range of strong interaction parameters.
If so, SQM can play an important
role in cosmology, neutron stars, cosmic ray physics, and relativistic
heavy-ion collisions. As an example of the intimate connections between
astrophysics and heavy-ion collision physics, this Chapter
gives an overview of the physical properties of
SQM in bulk and of small-baryon number strangelets; discusses
the possible formation, destruction, and implications of 
lumps of SQM (quark nuggets) 
in the early Universe; and describes the structure and signature of
strange stars, as well as the formation and detection of strangelets
in cosmic rays.
It is concluded, that astrophysical and laboratory searches are
complementary in many respects, and that both should be pursued to
test the intriguing possibility of a strange ground state for hadronic
matter, and (more generally) to improve our knowledge of the strong
interactions.
\end{abstract}

\section{Introduction}
\label{sec:intro}

Hadronic matter is expected to undergo a transition to quark-gluon
plasma under conditions of high temperature and/or baryon chemical
potential. These conditions may be achieved for a brief moment in
ultrarelativistic heavy-ion collisions, but they are also likely to
appear in Nature. A very high density (and comparatively low
temperature) environment exists in the interior of neutron stars, which
may actually contain significant amounts of quark matter in the
interior. High temperatures (but rather low baryon chemical potential)
were realized in the first $10^{-4}$ seconds after the Big Bang, and here
a hot quark-gluon plasma state must have existed until the temperature
dropped to 100--200 MeV due to the adiabatic expansion of the Universe.

This Chapter will outline some of the possible ways in which astrophysics
may teach us about the existence and properties of quark-gluon plasmas.
The advantage relative to laboratory searches is, that truly bulk
systems can be studied, and that the timescales involved are much longer
than those relevant to collisions. Disadvantages are that
astrophysicists (with possible exceptions if strange quark matter is
absolutely stable) can only observe indirect consequences of the plasma
state for example in the properties of pulsars or in the distribution of
light nuclei produced a few minutes after the Big Bang.
It will be shown, however, that astrophysics arguments in many cases can
be used to constrain parameters significantly relative to direct
experimental approaches because of the large volumes and timescales
involved.

The implications of quark-gluon plasmas in astrophysics and cosmology are
many-fold, and I shall focus
on aspects related to the idea of (meta)stability of strange quark
matter through discussions partly biased by my own research interests.

Lumps of up, down, and strange quarks (strange quark matter, SQM),
with masses ranging from small nuclei to neutron stars,
rather than $^{56}$Fe, could be
the ground state of hadronic matter even at zero temperature and pressure.
This possibility, first noted by Bodmer in 1971 \cite{bod71},
has attracted much attention since Witten resurrected the idea in 1984
\cite{wit84}.
The existence of stable or metastable SQM would have numerous
consequences for physics and astrophysics, and testing some of these
consequences should ultimately tell us whether SQM really exists.

First it was
believed that SQM might give a natural explanation of the cosmological
dark matter problem. While not ruled out, this idea is now
less popular, but strange quark matter may still be important in astrophysical
settings, such as strange stars.
Numerous investigations have searched for deposits of SQM on the Earth and
in meteorites, so far unsuccessfully, and recently relativistic heavy-ion
collision experiments have been performed and/or proposed to test the idea.
Cosmic ray searches have come up with a few potential candidates for small
SQM-lumps (strangelets), but at present no 
compelling evidence for stable SQM has been presented. This, however, does
not rule it out. Most searches for SQM are sensitive to strangelets with
very low baryon number, $A$, and as discussed later, 
finite size effects have a significant destabilizing effect
on such objects, even if SQM is stable in bulk.

There is a significant range of strong interaction parameters for which
SQM in bulk is stable. But even if it is not, many of the (astro)physical
implications are more or less unchanged in the case of metastable SQM.
In neutron stars, for instance, the high pressure brings SQM closer to
stability relative to hadronic matter, 
and it is quite likely, that neutron stars contain cores of
strange quark matter, even if SQM is unstable at zero pressure.
In relativistic heavy-ion collision experiments, strangelets need ``only''
survive for $10^{-8}$ seconds to be of interest.
In fact, (meta)stable strangelets may be one of the ``cleanest'' signatures
for formation of a quark-gluon plasma in such collisions.

The present review tries to give an account of the status of
strange quark matter physics and astrophysics, as of early 1998, but of
course not all aspects are covered in equal detail.
In particular, nothing is said about the heroic experimental efforts to
produce strangelets in heavy-ion collisions.
A collection of papers
describing all aspects of SQM and a list of references to the field
through mid-1991 can be found in \cite{sqm91}.
An earlier review was given in \cite{alcoli88}. Recent reviews
include \cite{kum94,kum95a,kum95b,mad94b,mad95a,mad95b,papal95,gresch98}, 
and the
thorough reader will notice, that some parts of the present Chapter
borrows from my own papers among these since the physics discussed
therein remains more or less unchanged. Refs.\ \cite{papal95,gresch98} also
discuss the related issue of lumps of metastable strange hadronic matter,
which will not be dealt with here.

Section \ref{sec:bulk}
discusses the physics of SQM, starting out with simple estimates of why
3-flavor quark-matter is likely to be more bound than the 2-flavor
alternative, proceeding with more detailed descriptions of SQM in bulk.
Smaller systems (strangelets), for which finite-size effects are
crucial, are described in Section \ref{sec:strangelet}. Most of the
results are based on the MIT bag-model, but it is worth stressing
from the outset that this should only be viewed as a crude approximation
to reality, ultimately to be surpassed by direct QCD-calculations.

Section \ref{sec:cosmo} deals with the possible production of lumps
of SQM (often called quark nuggets) in the cosmological quark-hadron phase
transition, and the struggle of quark nuggets to survive evaporation and
boiling in a hostile environment. It turns out, that only large nuggets
are likely to survive, but the physics involved in the destruction process
is illuminating as it resembles the (time-reversed)
physics involved in strangelet production
in heavy-ion collisions. Implications of surviving quark nuggets for
Big Bang nucleosynthesis and the dark matter problem are also discussed.

Perhaps the most likely place to discover SQM (even if it is not absolutely
stable) is in neutron stars. These could be ``hybrid'', ``strange'',
or even ``mixed''
(the first term conventionally used for neutron stars with quark cores;
the second for ``true'' quark stars in case of SQM stability, and the
latter for objects with mixed phases of quark matter and nuclear matter).
Section \ref{sec:star} describes these stars, their implications for our
understanding of pulsars, and the possible connection to the
energetic gamma-ray bursters.

Strangelets surviving from the early Universe {\it or\/} released from
strange stars in binary systems have been searched for in cosmic ray
detectors and in meteorites and mineral deposits.
So far there are only a few potential candidates, but more
sensitive experiments will soon be carried out. Section \ref{sec:cosmic}
discusses some of the limits obtained. It also
presents an astrophysical argument which either improves the Earth-based
flux-limits by many orders of magnitude (almost excluding absolutely
stable SQM), or predicts that all neutron stars
are strange stars, if SQM is stable (the prediction to choose depends on
whether any pulsars can be proven to be ordinary neutron stars). 

Conclusions and a brief outlook are provided in Section \ref{sec:concl}.

\section{Physics of SQM in bulk}
\label{sec:bulk}

\subsection{Does strange matter conflict with experience?}

At first sight, the possibility that quark matter could be absolutely stable
seems to contradict daily life experiences (and experiments) showing that
nuclei consist of neutrons and protons, rather than a soup of quarks. If a
lower energy state exists, then why are we here? Why have we not decayed
into strange quark matter?

The answer to this obvious question is, that (meta)stability of strange
quark matter requires a significant fraction of strange quarks to be present.
Conversion of an iron nucleus into an $A=56$ strangelet thus demands a very
high order weak interaction to change dozens of $u$- and $d$-quarks into $s$%
-quarks at the same time. Such a process has negligible probability of
happening. For lower $A$ the conversion requires a lower order weak
interaction, but as demonstrated later, finite-size effects destabilize
small strangelets so that they become unstable or only weakly metastable
even if strange quark matter is stable in bulk.

Therefore (meta)stability of strange quark matter does not conflict with the
existence of ordinary nuclei. On the other hand, the existence of ordinary
nuclei shows, that quark matter composed of $u$- and $d$-quarks alone is
unstable, a fact that will be used later on to place constraints on model
parameters.

Another constraint from our mere existence can be placed on the {\bf %
electrical charge} of strangelets. If energy is gained by converting
ordinary matter into strange quark matter, strangelets with negative quark
charge, even if globally neutral due to a cloud of positrons, would have
devastating consequences, eating up the nuclei they would encounter. 
Even a small stable
component in the cosmos would be intolerable (but they could still
appear as metastable products in heavy-ion collisions, like the recent
charge $-1$, mass 7.4 GeV event in NA52 at CERN \cite{kabal97}). 
A positive charge on the quark
surface (neutralized by surrounding electrons) is less problematic, because
ordinary nuclei will be electrostatically repelled. The barrier has to be of
a certain height, though, in order not to impact stellar evolution (see
below). Note that {\bf neutrons} are easily absorbed. As demonstrated later,
this has important consequences for quark star formation and can be used to
constrain strange matter properties using several astrophysical lines of
reasoning. It may even lead to practical applications in energy production,
etc.\ \cite{shaal89}.

\subsection{Simple arguments for (meta)stability}
\label{sub:simple}

As argued above, quark matter composed of $u$ and $d$-quarks is expected to
be unstable (except from 3-quark baryons). Introducing a third flavor makes
it possible to reduce the energy relative to a two-flavor system, because an
extra Fermi-well is available. The introduction of an extra fermion-flavor
makes it possible to increase the spatial concentration of quarks, thereby
reducing the total energy. A penalty is paid because the mass of the $s$%
-quark is high compared to that of $u$ and $d$, so stability is most likely
for low $s$-quark mass.

To make the argument slightly more quantitative, consider non-interacting,
massless quarks\footnote{%
Since current quark masses rather than constituent quark masses enter in the
MIT bag model used to describe SQM, this is a very good approximation for $u$
and $d$-quarks with $5\,$MeV$\approx m_u<m_d\approx 10\,{\rm MeV}\ll 
300\,{\rm MeV}
\approx \mu _u,\mu _d$.} inside a confining bag at temperature $T=0$,
without external pressure. For a massless quark-flavor, $i$, the Fermi
momentum, $p_{Fi}$, equals the chemical potential, $\mu _i$ (throughout the
chapter, unless otherwise noted, $\hbar =c=k_B=1$; for an introduction
to Fermi-gas thermodynamics, see for instance Ref.\ \cite{shateu83}). 
Thus the number
densities are $n_i=\mu _i^3/\pi ^2$, the energy densities $\epsilon _i=3\mu
_i^4/(4\pi ^2)$, and the pressures $P_i=\mu _i^4/(4\pi ^2)$. The sum of the
quark pressures is balanced by the confining bag pressure, $B$; $\sum_iP_i=B$; 
the total energy density is $\epsilon =\sum_i\epsilon _i+B=3\sum_iP_i+B=4B$, 
and the density of baryon number is $n_B=\sum_in_i/3$. Notice that the sum
of the constituents pressures, as well as the total energy density are given
solely in terms of the bag constant, $B$.

For a gas of $u$ and $d$-quarks charge neutrality requires $n_d=2n_u$, or 
$\mu _2\equiv \mu _u=2^{-1/3}\mu _d$. The corresponding two-flavor quark
pressure is $P_2=P_u+P_d=(1+2^{4/3})\mu _2^4/(4\pi ^2)=B$, the total energy
density $\epsilon _2=3P_2+B=4B$, and the baryon number density 
$n_{B2}=(n_u+n_d)/3=\mu _2^3/\pi ^2$, giving an energy per baryon of 
\begin{eqnarray}
\epsilon _2/n_{B2}=(1+2^{4/3})^{3/4}(4\pi
^2)^{1/4}B^{1/4}=6.441B^{1/4}\approx 934{\rm MeV}B_{145}^{1/4},
\end{eqnarray}
where $B_{145}^{1/4}\equiv B^{1/4}/145{\rm MeV}$; 145MeV being the lowest
possible choice for reasons discussed below.

A three-flavor quark gas is electrically neutral for $n_u=n_d=n_s$, i.\ e.\ 
$\mu _3\equiv \mu _u=\mu _d=\mu _s$. For fixed bag constant the three-quark
gas should exert the same pressure as the two-quark gas (leaving also the
energy density, $\epsilon _3=3P_3+B=4B$, unchanged). That happens when $\mu
_3=[(1+2^{4/3})/3]^{1/4}\mu _2$, giving a baryon number density of 
$n_{B3}=\mu _3^3/\pi ^2=[(1+2^{4/3})/3]^{3/4}n_{B2}$. The energy per baryon
is then 
\begin{eqnarray}
\epsilon _3/n_{B3}=3\mu _3=3^{3/4}(4\pi ^2)^{1/4}B^{1/4}=5.714B^{1/4}\approx
829{\rm MeV}B_{145}^{1/4};
\label{eps3}
\end{eqnarray}
{\it lower\/} than in the two-quark case by a factor 
$n_{B2}/n_{B3}=(3/(1+2^{4/3}))^{3/4}\approx 0.89$ .

The possible presence of electrons was neglected in the calculations above.
For two-flavor quark matter, including electrons in chemical equilibrium via 
$u+e^- \leftrightarrow d+\nu_e$, so that $\mu_u +\mu_e=\mu_d$, gives more
cumbersome equations, but only changes $\epsilon_2/n_{B2}$ to $6.445B^{1/4}$, 
since $\mu_e$ turns out to be rather small. Three-flavor quark matter does
not contain electrons for non-interacting, massless quarks.

One may therefore gain of order 100 MeV per baryon by introducing an extra
flavor. At fixed confining bag pressure the extra Fermi-well allows one to
pack the baryon number denser into the system, thereby gaining in binding
energy.

The energy per baryon in a free gas of neutrons is the neutron mass, 
$m_n=939.6{\rm MeV}$; in a gas of $^{56}$Fe it is 930 MeV. Naively, stability
of $ud$-quark matter relative to neutrons thus corresponds to $\epsilon
_2/n_{B2}<m_n$, or $B^{1/4}<145.9{\rm MeV}$ ($B^{1/4}<144.4{\rm MeV}$ for
stability relative to iron). The argument can be turned around:
Since one observes neutrons and $^{56}$Fe in Nature, rather than $ud$-quark
matter, it is concluded that $B^{1/4}$ must be larger than the numbers just
quoted. More detailed calculations including finite-size effects and
Coulomb-forces do not change these numbers much, so we shall assume for the
present purpose that $B^{1/4}=145$MeV is an experimental lower limit for 
$\alpha _s=0$. (Here $\alpha _s$ denotes the strong ``fine-structure''
constant; $\alpha _s=0$ corresponding to non-interacting quarks except
for the confinement given by $B$).

Bulk strange quark matter is absolutely stable relative to a gas of iron for 
$B^{1/4}<162.8 {\rm MeV}$, metastable relative to a neutron gas for 
$B^{1/4}<164.4 {\rm MeV}$, and relative to a gas of $\Lambda$-particles (the
ultimate production limit in heavy-ion collisions) for $B^{1/4}<195.2 {\rm 
MeV}$. These numbers are upper limits. As demonstrated below, a finite $s$%
-quark mass as well as a non-zero strong coupling constant decreases the
limit on $B^{1/4}$.

The presence of ordinary nuclei in Nature {\it cannot\/} be used to turn the
values of $B^{1/4}$ just quoted for SQM into lower limits. Conversion of a
nucleus into a lump of SQM requires simultaneous transformation of roughly 
$A $ $u$- and $d$-quarks into $s$-quarks. The probability for this to happen
involves a weak interaction coupling to the power $A$, i.\ e.\ it does not
happen. This leads to the conclusion, that even if SQM is the lowest energy
state for hadronic matter in bulk, its formation requires a strangeness-rich
environment or formation via a ``normal'' 
quark-gluon plasma in relativistic heavy-ion
collisions, the early Universe, or a neutron star interior. All of these
possibilities will be explored in the following.

\subsection{SQM in bulk at $T=0$}
\label{sub:bulksqm}

The estimates above assumed $m_s=\alpha _s=0$. Non-zero $\alpha _s$ was
found by Farhi and Jaffe \cite{farjaf84} to correspond effectively to a
reduction in $B$. In the interest of simplicity I will therefore set $\alpha
_s=0$ in most of the following. The energy ``penalty'' paid by having to
form $s$-quarks at a finite mass of 50--300 MeV calls for more detailed
calculations, however. Such calculations are usually performed within the
MIT bag model \cite{choal74,degal75}.

Strange quark matter contains degenerate Fermi gases of $u$, $d$, and $s$
quarks, and $e^{-}$ or $e^{+}$. Chemical equilibrium is maintained by weak
interactions, 
\begin{eqnarray}
d &\leftrightarrow &u+e^{-}+\bar{\nu}_e \\
s &\leftrightarrow &u+e^{-}+\bar{\nu}_e \\
u+s &\leftrightarrow &d+u,
\end{eqnarray}
where the first two reactions should be understood to include also the
various permutations of the involved particles.

Neutrinos generally escape the system, so we shall ascribe to them no
chemical potential. Thus the chemical potentials in equilibrium are given by 
\begin{equation}
\mu _d=\mu _s=\mu _u+\mu _e.  \label{chempot}
\end{equation}

Knowing the chemical potentials one can calculate the thermodynamic
potentials. 
\begin{eqnarray}
\Omega _{e,V} &=&-{\frac{{\mu _e^4}}{{12\pi ^2}}} \\
\Omega _{u,V} &=&-{\frac{{\mu _u^4}}{{4\pi ^2}}} \\
\Omega _{d,V} &=&-{\frac{{\mu _d^4}}{{4\pi ^2}}} \\
\Omega _{s,V} &=&-{\frac{{\mu _s^4}}{{4\pi ^2}}}\left( (1-\lambda
^2)^{1/2}(1-{\frac 52}\lambda ^2)+{\frac 32}\lambda ^4\ln {\frac{{%
1+(1-\lambda ^2)^{1/2}}}\lambda }\right) ,
\end{eqnarray}
defining $\lambda \equiv m_s/\mu _s$.

Number densities are given by 
\begin{equation}
n_{i,V}=-\partial \Omega _{i,V}/\partial \mu _i;
\end{equation}
i.\ e. $n_{e,V}=\mu _e^3/3\pi^2$, $n_{u,V}=\mu _u^3/\pi^2$, 
$n_{d,V}=\mu_d^3/\pi ^2$, $n_{s,V}=\mu_s^3(1-\lambda ^2)^{3/2}/\pi^2.$ 
The total pressure is 
\begin{equation}
P=\sum_iP_i-B=-\sum_i\Omega _{i,V}-B=0,  \label{presbal}
\end{equation}
and charge neutrality requires 
\begin{equation}
{\frac 23}n_{u,V}-{\frac 13}n_{d,V}-{\frac 13}n_{s,V}-n_{e,V}=0.
\label{neutral}
\end{equation}
The total energy density is 
\begin{equation}
\epsilon =\sum_i(\Omega _{i,V}+n_{i,V}\mu _i)+B,
\end{equation}
and the density of baryon number 
\begin{equation}
n_B={\frac 13}(n_{u,V}+n_{d,V}+n_{s,V}).
\end{equation}

\begin{figure}[h!tb]
\centerline{\psfig{file=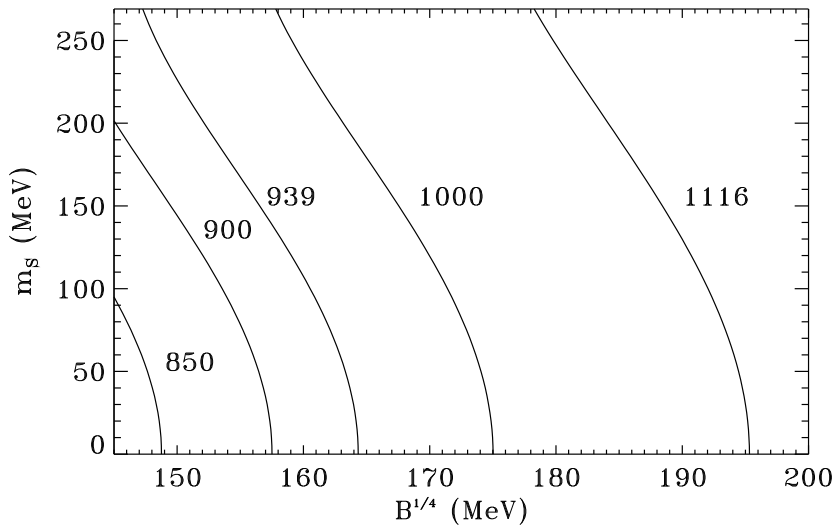,width=12cm}}
\caption{Energy per baryon for bulk strange quark matter as a function
of bag constant and strange quark mass.}
\label{fig:bul}
\end{figure}

Combining Eqs.~(\ref{chempot}) and (\ref{neutral}) leaves only one
independent chemical potential, which can be determined from the pressure
balance, Eq.~(\ref{presbal}). Thus all parameters can be calculated for a
given choice of $m_s$ and $B$. Results of such calculations are shown 
in Figure \ref{fig:bul}. Similar calculations were originally done by
Farhi and Jaffe \cite{farjaf84}.

The calculations above assumed zero temperature and external pressure.
Finite temperature and external pressure can be relevant in connection with
cosmology (Section \ref{sec:cosmo}) and strange stars (Section \ref{sec:star}) 
respectively, and also
for strangelet creation in collision experiments. The relevant extensions of
the formulae above will be given in Section \ref{subsec:fintem}.

\section{Strangelets}
\label{sec:strangelet}

So far the treatment of SQM has focused on the bulk properties. This
approximation is generally valid for large baryon numbers. For $A\ll 10^7$
the quark part of SQM is smaller than the Compton wavelength of electrons,
so electrons no longer ensure local charge neutrality. Therefore Coulomb
energy has to be taken into account, though the fortuitous cancellation of $%
q_u+q_d+q_s=\frac 23-\frac 13-\frac 13=0$ means that Coulomb energy is much
less important for strangelets than for nuclei. For even smaller baryon
numbers (in practice $A<10^3$) other finite size effects such as surface
tension and curvature have to be taken into account.

Several strangelet
searches with relativistic heavy-ion collisions as well as cosmic ray
searches have been carried out, and others are planned for the future. 
Most of these searches are sensitive only to low $A$-values, so it is important
to know the properties of small lumps of strange quark matter (strangelets).

In the following I will describe the physical properties of strangelets in
the language of the MIT-bag model (only limited work has been
performed using other models---qualitatively confirming the MIT-bag results,
though quantitative details can differ). First, I will discuss results
obtained from direct solution of the Dirac equation with MIT-bag boundary
conditions; such mode filling calculations correspond to a nuclear shell
model. Then I will show how the mean behavior of the shell model results can
be understood physically in terms of a liquid drop model calculation based
on a smoothed density of states, and how approximations to the liquid drop
results give simple formulae for strangelet masses etc. Finally I discuss
the changes introduced if strangelets are at finite rather than zero
temperature.

\subsection{Shell model}

Mode-filling for large numbers of quarks in a spherical MIT-bag was
performed for $ud$-systems by Vasak, Greiner and Neise \cite{vasgre86}, and
for 2- and 3-flavor systems by Farhi and Jaffe \cite{farjaf84}, and Greiner 
{\it et al.} \cite{greal88} (see also \cite{takboy88}). 
Gilson and Jaffe \cite{giljaf93} published an
investigation of low-mass strangelets for 4 different combinations
of $s$-quark mass and bag constant with particular emphasis on metastability
against strong decays. Further parameter ranges were studied and compared to
liquid drop model calculations by Madsen \cite{mad94}, and recently new
shell-model studies were published by Schaffner-Bielich {\it et al.}\
\cite{schgre97}. All of these
calculations were performed for $\alpha _s=0$, which will also be assumed in
the following.

In the MIT bag model noninteracting quarks are confined to a spherical
cavity of radius $R$. They satisfy the free Dirac equation inside the cavity
and obey a boundary condition at the surface, which corresponds to no
current flow across the surface. The bag itself has an energy of $BV$. In
the simplest version the energy (mass) of the system is given by the sum of
the bag energy and the energies of individual quarks, 
\begin{equation}
E=\sum_{i=u,d,s}\sum_\kappa 
N_{\kappa ,i}(m_i^2+k_{\kappa ,i}^2)^{1/2}+B4\pi R^3/3.
\label{bagenergy}
\end{equation}
Here $k_{\kappa ,i}\equiv x_{\kappa ,i}/R$, where $x_{\kappa ,i}$ are
eigenvalues of the equation 
\begin{equation}
f_\kappa (x_{\kappa ,i})={\frac{{-x_{\kappa ,i}}}{{(x_{\kappa
,i}^2+m_i^2R^2)^{1/2}+m_iR}}}f_{\kappa -1}(x_{\kappa ,i}).
\end{equation}
$f_\kappa $ are regular Bessel functions of order $\kappa $ , 
\begin{equation}
f_\kappa (x)=\left\{ 
\begin{array}{ll}
j_\kappa (x) & \kappa \geq 0 \\ 
y_\kappa (x)=(-1)^{\kappa +1}j_{-\kappa -1}(x) & \kappa <0
\end{array}
\right.
\end{equation}
For states with quantum numbers $(j,l)$ $\kappa $ takes the values $\kappa
=\pm (j+{\frac 12})$ for $l=j\pm {\frac 12}$. For a given quark flavor each
level has a degeneracy of $N_{\kappa, i}=3(2j+1)$ 
(the factor 3 from color degrees of
freedom). For example, the $1S_{1/2}$ ground-state ($j=1/2$, $l=0$, $\kappa
=-1$) for a massless quark corresponds to solving the equation $\tan
x=x/(1-x)$, giving $x\simeq 2.0428$. The ground state has a degeneracy of 6
per flavor.

For massless quarks (finding the equilibrium radius from $\partial
E/\partial R=0$) one gets 
\begin{equation}
E=364.00{\rm MeV}B_{145}^{1/4}\left( \sum x_{\kappa ,i}\right) ^{3/4}
\label{zerosum}
\end{equation}
where the sum is to be taken over all $3A$ quark-levels, and the numbers $%
x_{\kappa ,i}$ for massless quarks are tabulated in \cite{vasgre86}.

For massive quarks the level filling scheme is more cumbersome (see e.g.\
Refs.\ \cite{giljaf93,mad94}).  Fixing bag constant and quark-masses, for each
baryon number one must fill up the lowest energy levels for a choice of
radius; then vary the radius until a minimum energy is found ($\partial
E/\partial R=0$). Since levels cross, the order of levels is changing as a
function of $R$. This is easily seen in the Figures, where one notices
discontinuous changes in the position of shells.

\begin{figure}[h!tb]
\begin{tabular}{cc}
\psfig{file=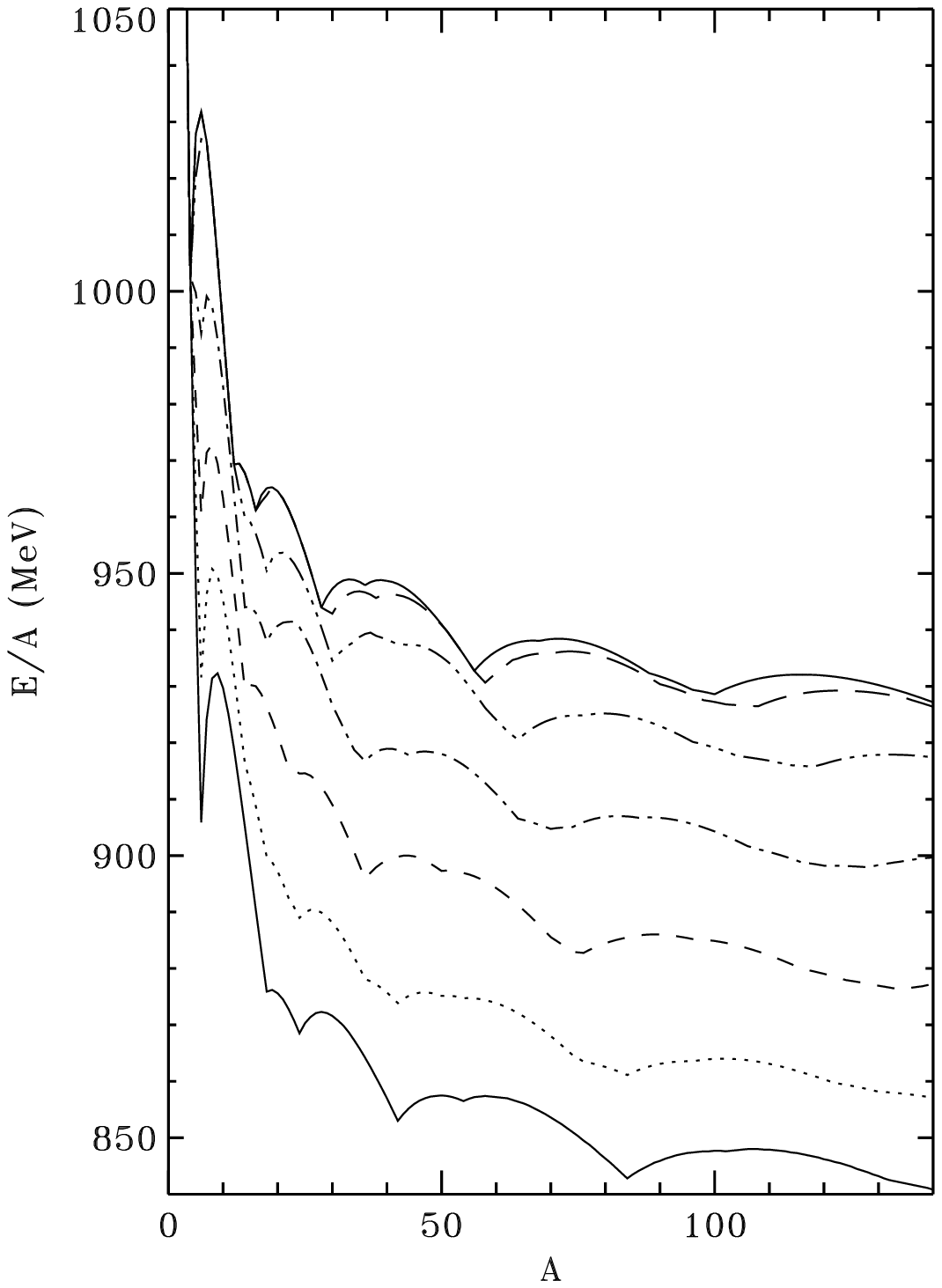,width=7.5cm}
\hspace{0cm}
\psfig{file=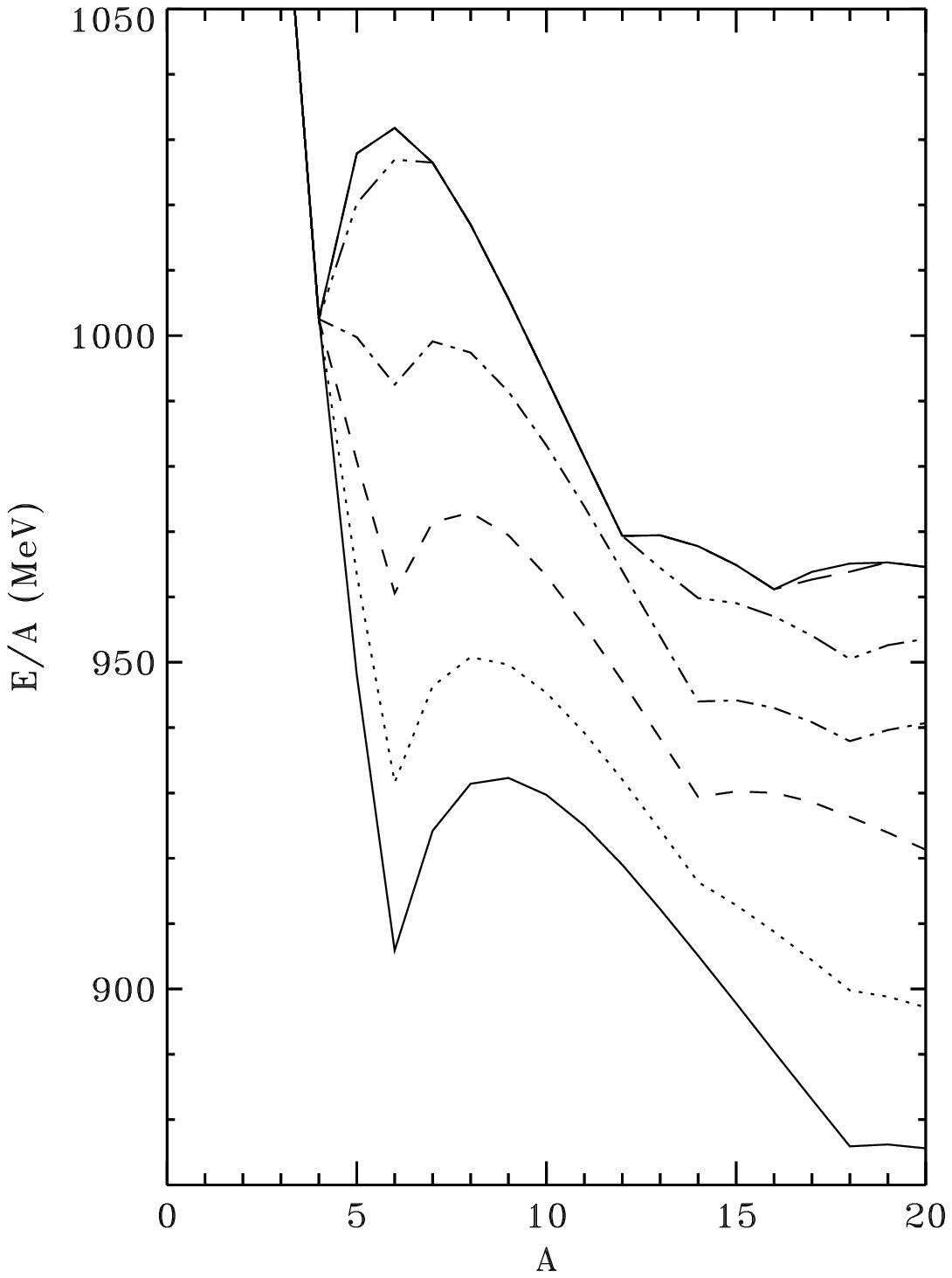,width=7.5cm}
\end{tabular}
\caption{Energy per baryon (in MeV) for strangelets with $B^{1/4}=145$MeV and
$m_s$ from 0--300 MeV in steps of 50 MeV ($m_s$ increases upward). The
figure on the right shows an expanded view of the low-mass region to
highlight the change of ``magic numbers'' with changing $m_s$.}
\label{fig:mode}
\end{figure}

One notices that the energy per baryon smoothly approaches the bulk limit
for $A\rightarrow \infty$, whereas the energy grows significantly for low $A$%
. For low $s$-quark mass shells are recognized for $A=6$ (3 colors and 2
spin orientations per flavor), and less conspicuous ones for $A=18$, 24, 42,
54, 60, 84, 102 etc. As $m_s$ increases it becomes more and more favorable
to use $u$ and $d$ rather than $s$-quarks, and the ``magic numbers'' change;
for instance the first closed shell is seen for $A=4$ rather than 6.

Equation (\ref{bagenergy}) can be modified by inclusion of Coulomb energy
and zero-point fluctuation energy. As already discussed the Coulomb energy
is generally small. The zero-point energy is normally included as a
phenomenological term of the form $-Z_0/R$, where fits to light hadron
spectra indicate the choice $Z_0=1.84$. This was used, for instance, by
Gilson and Jaffe \cite{giljaf93}. Roughly half of this phenomenological term
is due to center-of-mass motion, which can be included more explicitly by
substituting $\left[ \left( \sum x_{\kappa ,i}\right) ^2-\sum x_{\kappa
,i}^2\right] ^{3/8}$ instead of $\left( \sum x_{\kappa ,i}\right) ^{3/4}$ in
Eq.~(\ref{zerosum}). The proper choice of $\alpha _s$ and $Z_0$ is a tricky
question. As discussed by Farhi and Jaffe \cite{farjaf84} the values are
intimately coupled to $B$ and $m_s$, and it is not obvious that values
deduced from bag model fits to ordinary hadrons are to be preferred. This
uncertainty may have an important effect for $A<5$--$10$, but the
zero-point energy quickly becomes negligible for increasing $A$ for reasons
explained in Section \ref{subsub:msbulk}. It means, however, 
that it is difficult to match
strangelet calculations to experimental data concerning ordinary hadrons or
limits on the putative $A=2$ $H$-dibaryon.

\subsection{Liquid drop model}

Mode-filling calculations are rather tedious but do of course give the
``correct'' results as far as the model can be trusted. But for many
applications a global mass-formula analogous to the liquid drop model for
nuclei is of great use and also gives further physical insight.

A phenomenological approach to a strangelet mass-formula was undertaken by
Crawford {\it et al.}\ \cite{descra93,crades93} 
whereas Berger and Jaffe \cite{berjaf87} made a
detailed analysis within the MIT bag model. They included Coulomb
corrections and surface tension effects stemming from the depletion in the
surface density of states due to the mass of the strange quark. Both effects
were treated as perturbations added to a bulk solution with the surface
contribution derived from a multiple reflection expansion. Madsen \cite
{mad94,mad93a,mad93b} gave a self-consistent treatment including also the
very important curvature energy.

The following discussion closely follows \cite{mad94}. All calculations
are done for zero temperature and strong coupling constant, $\alpha _s$. As
argued by Farhi and Jaffe \cite{farjaf84} the latter assumption can be
relaxed by a re-scaling of the bag constant. Also, I shall concentrate on
systems small enough ($A<10^7$) to justify neglect of electrons. Strangelets
with $A\ll 10^7$ are smaller than the electron Compton wavelength, and
electrons are therefore mainly localized outside the quark phase.  Thus
strangelets do not obey a requirement of local charge neutrality, as was the
case for SQM in bulk. This leads to a small Coulomb energy, which is
rather negligible for the mass-formula (less than a few MeV per baryon), but
which is decisive for the charge-to-mass ratio of the strangelet. A
characteristic of strangelets, which is perhaps the best experimental
signature, is that this ratio is very small compared to ordinary nuclei.
Finally, I neglect charge screening, an issue of negligible importance for
the mass formula, but of some importance for the charge-to-mass ratio for
systems of radii above 5--10 fm ($A>10^2$--$10^3$) \cite{hei93}.

In the ideal Fermi-gas approximation the energy of a system composed of
quark flavors $i$ is given by 
\begin{equation}
E=\sum_i(\Omega_i+N_i\mu_i)+BV+E_{{\rm Coul}}.  \label{Estrangelet2}
\end{equation}
Here $\Omega_i$, $N_i$ and $\mu_i$ denote thermodynamic potentials, total
number of quarks, and chemical potentials, respectively. $B$ is the bag
constant, $V$ is the bag volume, and $E_{{\rm Coul}}$ is the Coulomb energy.

In the multiple reflection expansion framework of Balian and Bloch \cite
{balblo70}, the thermodynamical quantities can be derived from a density of
states of the form 
\begin{equation}
{\frac{{dN_i}}{{dk}}}=6\left\{ {\frac{{k^2V}}{{2\pi ^2}}}+f_S\left( {\frac{%
m_i}k}\right) kS+f_C\left( {\frac{m_i}k}\right) C+....\right\} ,
\label{dNdk}
\end{equation}
where area $S=\oint dS$ ($=4\pi R^2$ for a sphere) and extrinsic curvature $%
C=\oint \left( {\frac 1{{R_1}}}+{\frac 1{{R_2}}}\right) dS$ ($=8\pi R$ for a
sphere). Curvature radii are denoted $R_1$ and $R_2$. For a spherical system 
$R_1=R_2=R$. The functions $f_S$ and $f_C$ will be discussed below.

In terms of volume-, surface-, and curvature-densities, $n_{i,V}$, $n_{i,S}$%
, and $n_{i,C}$, the number of quarks of flavor $i$ is 
\begin{equation}
N_i=\int_0^{k_{Fi}}{\frac{{dN_i}}{{dk}}}dk=n_{i,V}V+n_{i,S}S+n_{i,C}C,
\end{equation}
with Fermi momentum $k_{Fi}=(\mu_i^2-m_i^2)^{1/2}=\mu_i(1-\lambda_i^2)^{1/2}$%
; $\lambda_i\equiv m_i/\mu_i$.

The corresponding thermodynamic potentials are related by 
\begin{equation}
\Omega _i=\Omega _{i,V}V+\Omega _{i,S}S+\Omega _{i,C}C,
\end{equation}
where $\partial \Omega _i/\partial \mu _i=-N_i$, and $\partial \Omega
_{i,j}/\partial \mu _i=-n_{i,j}$. The volume terms are given by 
\begin{eqnarray}
\Omega _{i,V}=-{\frac{{\mu _i^4}}{{4\pi ^2}}}\left[ (1-\lambda _i^2)^{1/2}(1-%
{\frac 52}\lambda _i^2)+{\frac 32}\lambda _i^4\ln {\frac{{1+(1-\lambda
_i^2)^{1/2}}}{\lambda _i}}\right] ,
\end{eqnarray}
\begin{equation}
n_{i,V}={\frac{{\mu _i^3}}{{\pi ^2}}}(1-\lambda _i^2)^{3/2}.
\end{equation}

The surface contribution from massive quarks is derived from
\begin{equation}
f_S\left({\frac mk}\right) =-{\frac 1{8\pi}}\left\{ 1-\left({\frac {2}{\pi}}
\right) \tan ^{-1}{\frac km}\right\} 
\end{equation}
as \cite{berjaf87} 
\begin{eqnarray}
\Omega _{i,S} &=&{\frac 3{{4\pi }}}\mu _i^3\left[ {\frac{{(1-\lambda _i^2)}}6%
}-{\frac{{\lambda _i^2(1-\lambda _i)}}3}\right.  \\
&&\left. -{\frac 1{{3\pi }}}\left( \tan ^{-1}\left[ {\frac{{(1-\lambda
_i^2)^{1/2}}}{\lambda _i}}\right] -2\lambda _i(1-\lambda _i^2)^{1/2}+\lambda
_i^3\ln \left[ {\frac{{1+(1-\lambda _i^2)^{1/2}}}{\lambda _i}}\right]
\right) \right] ;  \nonumber
\end{eqnarray}
\begin{eqnarray}
n_{i,S}=-{\frac 3{{4\pi }}}\mu _i^2\left[ {\frac{{(1-\lambda _i^2)}}2}-{%
\frac 1{{\pi }}}\left( \tan ^{-1}\left[ {\frac{{(1-\lambda _i^2)^{1/2}}}{%
\lambda _i}}\right] -\lambda _i(1-\lambda _i^2)^{1/2}\right) \right] .
\end{eqnarray}
For massless quarks $\Omega _{i,S}=n_{i,S}=0$, whereas $f_C(0)=-1/24\pi ^2$
gives \cite{farjaf84,mad93a,mad93b} $\Omega _{i,C}={\mu _i^2}/8\pi ^2$; 
$n_{i,C}=-{ \mu _i}/4\pi ^2$.

The curvature terms have never been derived for massive quarks, but as shown
by Madsen \cite{mad94}, the following {\it Ansatz\/} (found from analogies
with the surface term and other known cases) works: 
\begin{equation}
f_C\left( {\frac mk}\right) ={\frac 1{{12\pi ^2}}}\left\{ 1-{\frac 32}{\frac
km}\left( {\frac \pi 2}-\tan ^{-1}{\frac km}\right) \right\} .
\end{equation}
This expression has the right limit for massless quarks ($f_C=-1/24\pi ^2$)
and for infinite mass, which corresponds to the Dirichlet boundary
conditions studied by Balian and Bloch \cite{balblo70} ($f_C=1/12\pi ^2$).
Furthermore, the expression gives perfect fits to mode-filling calculations
(see the Figures and discussion below). From this {\it Ansatz\/} one derives
the following thermodynamical potential and density: 
\begin{eqnarray}
\Omega _{i,C}={\frac{{\mu _i^2}}{{8\pi ^2}}}\left[ \lambda _i^2\log {\frac{{%
1+(1-\lambda _i^2)^{1/2}}}{{\lambda _i}}}+{\frac \pi {{2\lambda _i}}}-{\frac{%
{3\pi \lambda _i}}{{2}}}+\pi \lambda _i^2-{\frac 1{\lambda _i}}\tan ^{-1}{%
\frac{{(1-\lambda _i^2)^{1/2}}}{{\lambda _i}}}\right] ;
\end{eqnarray}
\begin{eqnarray}
n_{i,C}={\frac{\mu _i}{{8\pi ^2}}}\left[ (1-\lambda _i^2)^{1/2}-{\frac{{3\pi 
}}2}{\frac{{(1-\lambda _i^2)}}{\lambda _i}}+{\frac 3{\lambda _i}}\tan ^{-1}{%
\frac{{(1-\lambda _i^2)^{1/2}}}{\lambda _i}}\right] .
\end{eqnarray}

With these prescriptions the differential of $E(V,S,C,N_i)$ is given by 
\begin{eqnarray}
dE=\sum_i\left( \Omega _{i,V}dV+\Omega _{i,S}dS+\Omega _{i,C}dC+\mu
_idN_i\right) +BdV+dE_{{\rm Coul}}.  \label{dE}
\end{eqnarray}

Minimizing the total energy at fixed $N_i$ by taking $dE=0$ for a sphere
gives the pressure equilibrium constraint 
\begin{equation}
B=-\sum_i\Omega _{i,V}-{\frac 2R}\sum_i\Omega _{i,S}-{\frac 2{R^2}}%
\sum_i\Omega _{i,C}-{\frac{{dE_{{\rm Coul}}}}{{dV}}},  \label{bag}
\end{equation}
with 
\begin{equation}
E_{{\rm Coul}}={\frac{{\alpha Z_V^2}}{{10R}}}+{\frac{{\alpha Z^2}}{{2R}}},
\end{equation}
\begin{equation}
{\frac{{dE_{{\rm Coul}}}}{{dV}}}=-{\frac{{\alpha Z_V^2}}{{40\pi R^4}}}-{%
\frac{{\alpha Z^2}}{{8\pi R^4}}},
\end{equation}
where $Z_V=\sum_iq_in_{i,V}V$ is the volume part of the total charge, $Z$,
whereas charge $Z-Z_V=\sum_iq_i(n_{i,S}S+n_{i,C}C)$ is distributed on the
surface. The quark charges are $q_u=2/3$, $q_d=q_s=-1/3$. Eliminating $B$
from Eq.~(\ref{Estrangelet2}) then gives the energy for a spherical quark
lump as 
\begin{equation}
E=\sum_i(N_i\mu _i+{\frac 13}\Omega _{i,S}S+{\frac 23}\Omega _{i,C}C)+{\frac
43}E_{{\rm Coul}}.  \label{Estrangelet3}
\end{equation}

The optimal composition for fixed baryon number, $A$, can be found by
minimizing the energy with respect to $N_i$ at fixed $V$, $S$, and $C$
giving 
\begin{equation}
0=dE=\sum_i\left( \mu _i+{\frac{{\partial E_{{\rm Coul}}}}{{\partial N_i}}}%
\right) dN_i.  \label{compo}
\end{equation}

\subsubsection{Massless quarks---Bulk limit}

For uncharged bulk quark matter Eq.~(\ref{Estrangelet3}) reduces to the
usual result for the energy per baryon 
\begin{equation}
\epsilon ^0=A^{-1}\sum_iN_i^0\mu _i^0,
\end{equation}
where superscript $0$ denotes bulk values. The energy minimization, Eq.~(\ref
{bag}), corresponds to 
\begin{equation}
B=-\sum_i\Omega _{i,V}^0=\sum_i{\frac{{(\mu _i^0)^4}}{{4\pi ^2}}}.
\label{bagmin}
\end{equation}
The last equality assumes massless quarks. In the bulk limit the baryon
number density is given by 
\begin{equation}
n_A^0={\frac 13}\sum_i{\frac{{(\mu _i^0)^3}}{{\pi ^2}}},
\end{equation}
and one may define a bulk radius per baryon as 
\begin{equation}
R^0=(3/4\pi n_A^0)^{1/3}.  \label{bulkrad}
\end{equation}

For quark matter composed of massless $u$, $d$, and $s$-quarks, the Coulomb
energy vanishes at equal number densities due to the fact that the sum of
the quark charges is zero. Thus it is energetically most favorable to have
equal chemical potentials for the three flavors. From the equations above
one may derive the following bulk expressions for 3-flavor quark matter: 
\begin{equation}
\mu _i^0=\left( {\frac{{4\pi ^2B}}3}\right) ^{1/4}=1.905B^{1/4}=276.2{\rm MeV%
}B_{145}^{1/4};
\end{equation}
\begin{equation}
n_A^0=(\mu _i^0)^3/\pi ^2=0.700B^{3/4}
\end{equation}
\begin{equation}
R^0=(3/4\pi n_A^0)^{1/3}=0.699B^{-1/4}.
\end{equation}
And the energy per baryon is 
\begin{equation}
\epsilon ^0=3\mu _i^0=5.714B^{1/4},
\end{equation}
in agreement with Eq.\ (\ref{eps3}).

Following Berger and Jaffe \cite{berjaf87} one may to first order regard
Coulomb, surface (and here correspondingly curvature) energies as
perturbations on top of the bulk solution. In this approach one gets 
\begin{eqnarray}
{\frac EA} &=&\epsilon ^0+A^{-1}\sum_i\Omega _{i,C}^0C^0=\epsilon ^0+{\frac{{%
3^{13/12}B^{1/4}}}{{\pi ^{1/6}2^{1/6}A^{2/3}}}}  \nonumber  \\
&\approx &\left[ 829{\rm MeV}+351{\rm MeV}A^{-2/3}\right] B_{145}^{1/4}.
\label{EoverA}
\end{eqnarray}
The corresponding result for 2-flavor quark matter (c.f.~\cite{mad93b}) is 
\begin{eqnarray}
{\frac EA}=\epsilon ^0+A^{-1}\sum_i\Omega _{i,C}^0C^0\approx \left[ 934{\rm %
MeV}+291{\rm MeV}A^{-2/3}\right] B_{145}^{1/4}.
\label{twoflavor}
\end{eqnarray}

\subsubsection{Massive s-quarks---Bulk limit}
\label{subsub:msbulk}

For $m_s>0$ the energy minimization, Eq.~(\ref{bagmin}), changes to 
\begin{eqnarray}
B &=&-\sum_i\Omega _{i,V}^0  \nonumber \\
&=&\sum_{i=u,d}{\frac{{(\mu _i^0)^4}}{{4\pi ^2}}}+{\frac{{(\mu _s^0)^4}}{{%
4\pi ^2}}}\left[ (1-\lambda ^2)^{1/2}(1-{\frac 52}\lambda ^2)+{\frac 32}%
\lambda ^4\ln {\frac{{1+(1-\lambda ^2)^{1/2}}}\lambda }\right] ,
\end{eqnarray}
and the baryon number density is now given by 
\begin{equation}
n_A^0={\frac 13}\left[ \sum_{i=u,d}{\frac{{(\mu _i^0)^3}}{{\pi ^2}}}+{\frac{{%
(\mu _s^0)^3}}{{\pi ^2}}}(1-\lambda ^2)^{3/2}\right] .
\end{equation}
A bulk radius per baryon is still defined by Eq.~(\ref{bulkrad}).

In bulk equilibrium the chemical potentials of the three quark flavors are
equal, $\mu _u^0=\mu _d^0=\mu _s^0\equiv \mu ^0=\epsilon ^0/3$. Neglecting
Coulomb energy one may approximate the energy per baryon of small
strangelets as a sum of bulk, surface and curvature terms, using the
chemical potential calculated in bulk: 
\begin{equation}
{\frac EA}=\epsilon ^0+A^{-1}\sum_i\Omega _{i,S}^0S^0+A^{-1}\sum_i\Omega
_{i,C}^0C^0,
\end{equation}
where $S^0=4\pi (R^0)^2A^{2/3}$ and $C^0=8\pi (R^0)A^{1/3}$. Examples for $%
B^{1/4}=145{\rm MeV}$ are (with $s$-quark mass in MeV given in parenthesis)

\begin{eqnarray}
\epsilon (0) &=&829{\rm MeV}+0{\rm MeV}A^{-1/3}+351{\rm MeV}A^{-2/3} \\
\epsilon (50) &=&835{\rm MeV}+61{\rm MeV}A^{-1/3}+277{\rm MeV}A^{-2/3} \\
\epsilon (150) &=&874{\rm MeV}+77{\rm MeV}A^{-1/3}+232{\rm MeV}A^{-2/3} \\
\epsilon (200) &=&896{\rm MeV}+53{\rm MeV}A^{-1/3}+242{\rm MeV}A^{-2/3} \\
\epsilon (250) &=&911{\rm MeV}+22{\rm MeV}A^{-1/3}+266{\rm MeV}A^{-2/3} \\
\epsilon (300) &=&917{\rm MeV}+0.3{\rm MeV}A^{-1/3}+295{\rm MeV}A^{-2/3} \\
\epsilon (350) &=&917{\rm MeV}+0{\rm MeV}A^{-1/3}+296{\rm MeV}A^{-2/3}
\label{udmass}
\end{eqnarray}

\begin{figure}[h!tb]
\centerline{\psfig{file=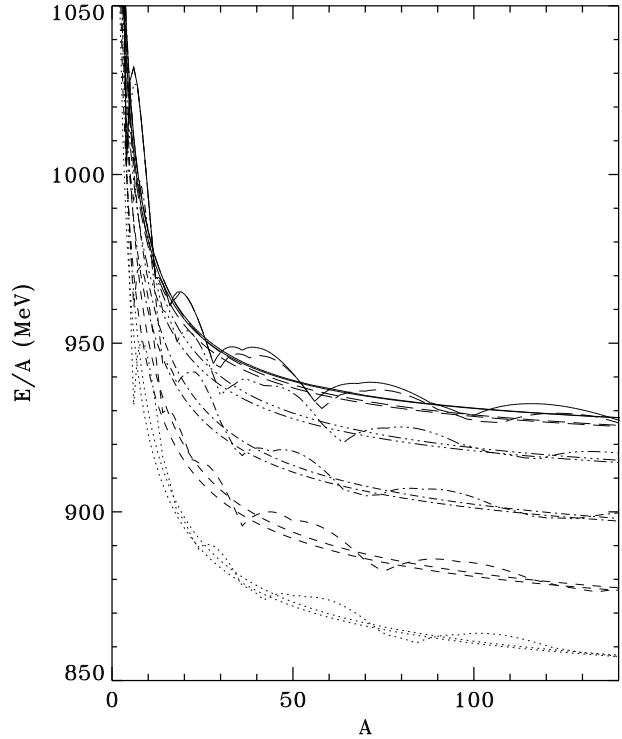,width=9cm}}
\caption{Shell-model and liquid drop model results compared for
$B^{1/4}=145$MeV with massless $u$ and $d$ quarks, and with $m_s$ in the
range 50--300 MeV in steps of 50 MeV. For each value of $m_s$ the upper
smooth curve is the full liquid drop model result, whereas the lower
smooth curve is the bulk approximation.}
\label{fig:bulkfit}
\end{figure}

The bulk approximations above generally undershoot the correct solution with
properly smoothed density of states by 2MeV for $A>100$, 5MeV for $A\approx 50$,
10MeV for $A\approx 10$ and 20MeV for $ A\approx 5$ (Figure \ref{fig:bulkfit}).
This is because the actual chemical potentials of the quarks
increase when $A$ decreases, whereas the bulk approximations use constant $%
\mu $. For massless $s$-quarks the expression for $\epsilon (0)$ scales
simply as $B^{1/4}$. The same scaling applies for $m_s>\epsilon ^0/3$, where
no $s$-quarks are present; in the example above the scaling can be applied
to $\epsilon (350)$. For intermediate $s$-quark masses both $\epsilon $ and $%
m_s$ should be multiplied by $B_{145}^{1/4}$ to scale the results. For
instance, if $B^{1/4}=165$MeV one finds $\epsilon (150)=985{\rm MeV}+93{\rm %
MeV}A^{-1/3}+265{\rm MeV}A^{-2/3}$; $\epsilon (250)=1027{\rm MeV}+46{\rm MeV}%
A^{-1/3}+284{\rm MeV}A^{-2/3}$. Coulomb effects were not included above.
Their inclusion would have no influence for $m_s\rightarrow 0$, but would
change the results by a few MeV for large $m_s$. In particular charge
neutral $ud$-quark matter has $\epsilon =\left[ 934{\rm MeV}+291{\rm MeV}%
A^{-2/3}\right] B_{145}^{1/4}$ (Eq.\ (\ref{twoflavor})) 
rather than the $\left[ 917{\rm 
MeV}+296{\rm MeV}A^{-2/3}\right] B_{145}^{1/4}$ found above (Eq. (\ref
{udmass})).

In connection with the shell-model calculations I described the effects of a
zero-point energy of the form $-Z_0/R$, and claimed that it was important
only for $A<10$. This can be understood in the bulk approximation of
constant $\mu $, because the zero-point term per baryon is proportional to $%
A^{-4/3}$ compared to $A^{-1/3}$ and $A^{-2/3}$ for surface and curvature
energies. The full term to be added to the bulk approximation expressions
for a given $\epsilon ^0$ is: 
\begin{equation}
\epsilon _{{\rm zero}}=-Z_0(4/243\pi )^{1/3}\left[ 2+[1-(3m_s/\epsilon
^0)]^{3/2}\right] ^{1/3}\epsilon ^0A^{-4/3},
\end{equation}
typically of order $-200Z_0{\rm MeV}A^{-4/3}$.

\subsection{Shell model versus liquid drop model}

Self-consistent solutions can be obtained from Eq.~(\ref{Estrangelet3}).
These solutions are compared to the shell-model calculations and the bulk
approximations in the Figures. The fits are very good, showing that
inclusion of surface tension and curvature energy via the multiple
reflection expansion explains the overall behavior of the results.

\begin{figure}[h!tb]
\begin{tabular}{cc}
\psfig{file=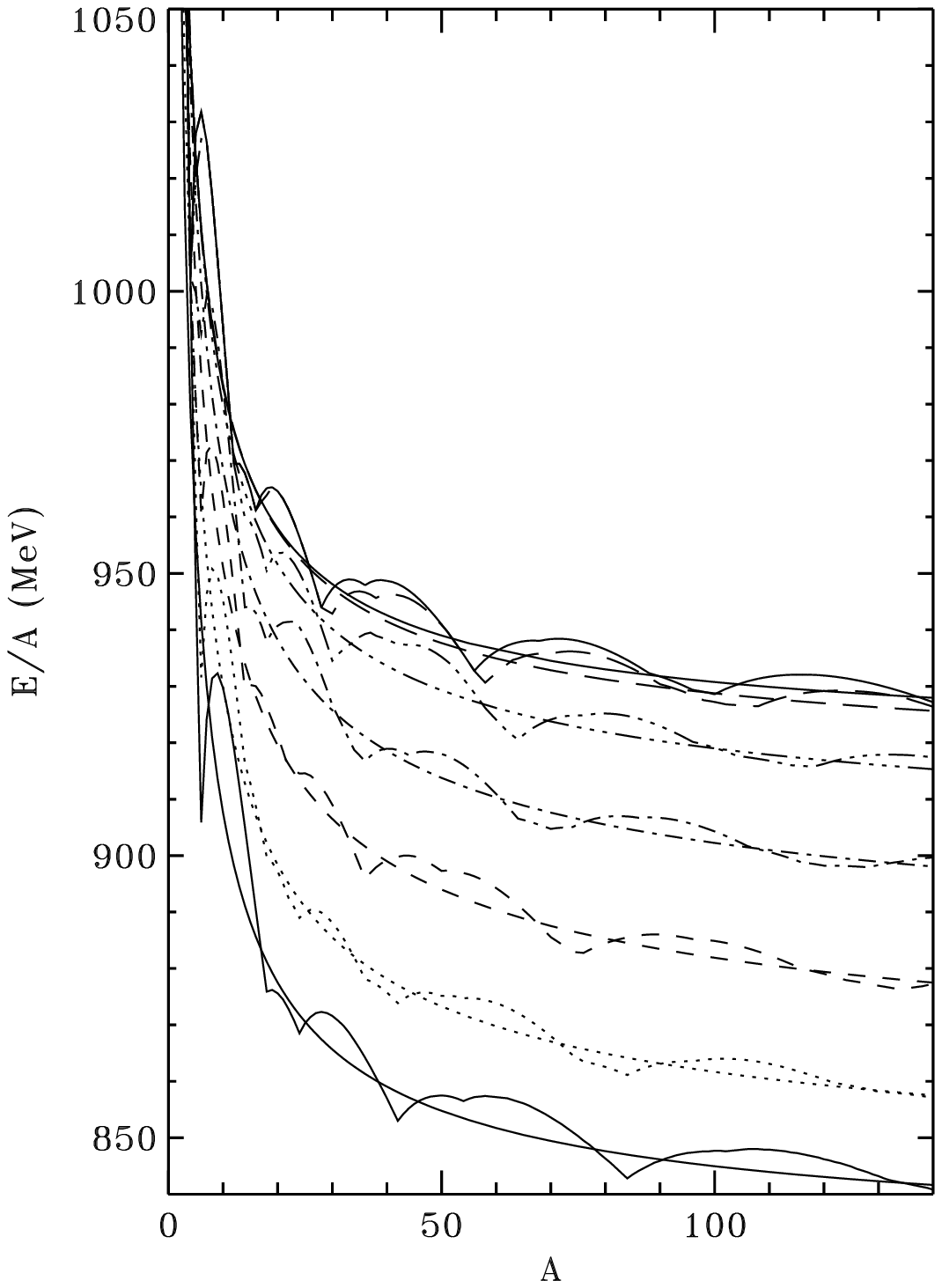,width=7.5cm}
\hspace{0cm}
\psfig{file=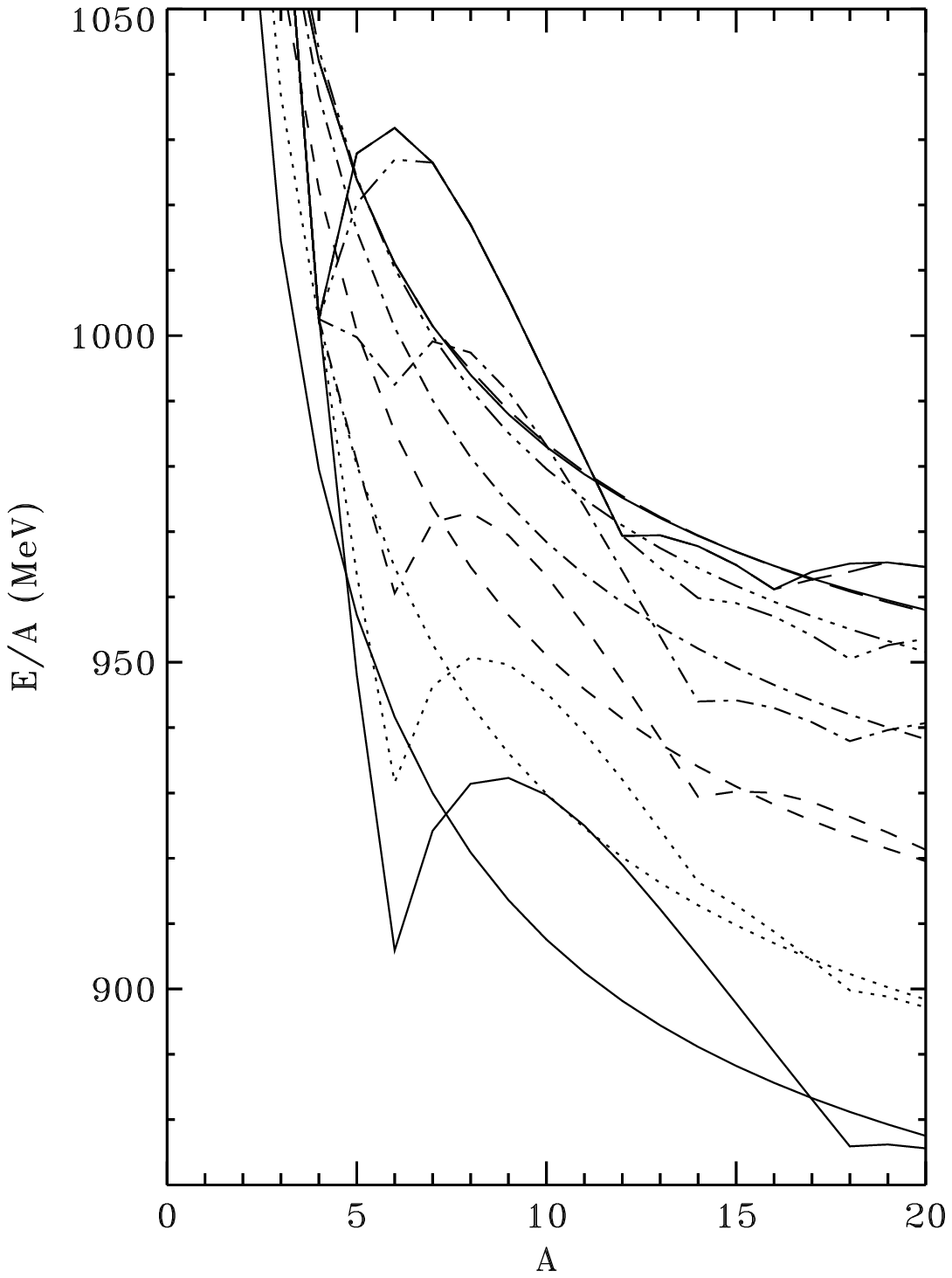,width=7.5cm}
\end{tabular}
\caption{As Figure \protect{\ref{fig:mode}} but showing also the liquid drop
results.}
\label{fig:liq}
\end{figure}

3-flavor quark matter is energetically favored in bulk, and could be
absolutely stable relative to $^{56}$Fe for $144 {\rm MeV}<B^{1/4}<163 {\rm %
MeV}$. The lower limit corresponds to experimentally excluded stability of $%
ud$ quark matter, whereas the upper limit corresponds to a bulk energy per
baryon of $uds$-matter of 930 MeV for $m_s=0$.

Finite-size systems are strongly destabilized by
the curvature energy, with a magnitude of about $300{\rm MeV}%
A^{-2/3}B_{145}^{1/4}$ for 3 quark flavors. This may pose problems for the
experimental attempts of producing strange quark matter, since these
experiments so far can only hope to create quark lumps with baryon number $%
A<20$--$30$, and observe lifetimes exceeding $10^{-8}$ seconds. Further
destabilization occurs for finite-mass $s$-quarks, where the surface tension
(exactly zero for massless quarks) adds up to $90{\rm MeV}A^{-1/3}$ to the
energy.

Writing $E/A=\epsilon ^0+c_{{\rm surf}}A^{-1/3}+c_{{\rm curv}}A^{-2/3}$,
with $c_{{\rm surf}}\approx 100$MeV and $c_{{\rm curv}}\approx 300$MeV, the
stability condition $E/A<m_n$ may be written as $A>A_{{\rm min}}^{{\rm abs}}$%
, where 
\begin{equation}
A_{{\rm min}}^{{\rm abs}}=\left( {\frac{{c_{{\rm surf}}+[c_{{\rm surf}%
}^2+4c_{{\rm curv}}(m_n-\epsilon ^0)]^{1/2}}}{{2(m_n-\epsilon ^0)}}}\right)
^3.
\end{equation}
Stability at baryon number 30 requires a bulk binding energy in excess of 65
MeV, which is barely within reach for $m_s>100$MeV if, at the same time, $ud$%
-quark matter shall be unstable. The proposed cosmic ray
strangelet-candidates with baryon number 370 \cite{saial90} would for
stability require a bulk binding energy per baryon exceeding 20~MeV to
overcome the combined curvature and surface energies. Absolute stability
relative to a gas of $^{56}$Fe corresponds to furthermore using 930~MeV
instead of $m_n$, whereas stability relative to a gas of $\Lambda $%
-particles (the ultimate limit for formation of short-lived strangelets)
would correspond to substitution of $m_\Lambda =1116$MeV.

Another way of stating the results is to calculate the minimum baryon number
for which long-lived metastability with respect to neutron emission is
possible. This requires $dE_{{\rm curv}}/dA+dE_{{\rm surf}}/dA<m_n-\epsilon
^0$, or 
\begin{equation}
A_{{\rm \min }}^{{\rm meta}}=\left( {\frac{{c_{{\rm surf}}+[c_{{\rm surf}%
}^2+3c_{{\rm curv}}(m_n-\epsilon ^0)]^{1/2}}}{{3(m_n-\epsilon ^0)}}}\right)
^3.
\end{equation}
To have $A_{{\rm min}}^{{\rm meta}}<30$ requires $m_n-\epsilon ^0>30$MeV,
which is possible, but only for a narrow range of parameters.

This should not, however, defer experimentalists from pursuing the proposed
searches. After all, the MIT bag model is only an approximation, and in
particular shell effects can have a stabilizing effect. As stressed by
Gilson and Jaffe \cite{giljaf93} the fact that the slope of $E/A$ versus $A$
becomes very steep near magic numbers can lead to strangelets that are
metastable (stable against single baryon emission) even for $\epsilon ^0>930$%
MeV. Also, the time-scale for energetically allowed decays has not been
calculated. Pauli-blocking is known to delay weak quark conversion in
strangelets \cite{heimad86,koc91,mad93c,hei92}, 
and this will probably have a significant influence on the
lifetimes. The existence of small baryon number strangelets is ultimately
an experimental issue.

\subsection{Strangelets at finite temperature}
\label{subsec:fintem}

Whereas the calculations above deal with strangelets at zero temperature,
the environment in heavy ion collisions is expected to be hot. An advantage
of the asymptotic mass formula compared to the shell-model calculations is,
that it can fairly easily be generalized to non-zero temperature.

The general expression for the thermodynamic potential, $\Omega _i$, is 
\begin{equation}
\Omega _i=\mp g_iT\int_0^\infty dk\frac{dN_i}{dk}\ln \left[ 1\pm \exp
(-(\epsilon (k)-\mu )/T)\right]
\end{equation}
where the upper sign is for fermions, the lower for bosons, and the density
of states, $\frac{dN_i}{dk}$, is given by Eq.~(\ref{dNdk}). For {\it %
massless\/} quarks (including antiquarks) an integration gives, per flavor, 
\begin{eqnarray}
\Omega _q=-\left( {\frac{{7\pi ^2}}{{60}}}T^4+{\frac{{\mu ^2T^2}}2}+{\frac{{%
\mu ^4}}{{4\pi ^2}}}\right) V+\left( {\frac{{T^2}}{{24}}}+{\frac{{\mu ^2}}{{%
8\pi ^2}}}\right) C,
\end{eqnarray}
with a corresponding quark number 
\begin{eqnarray}
N_q=-{\frac{{\partial \Omega _q}}{{\partial \mu }}}=\left( \mu T^2+{\frac{{%
\mu ^3}}{{\pi ^2}}}\right) V-{\frac{{\mu }}{{4\pi ^2}}}C.
\end{eqnarray}
For gluons 
\begin{equation}
\Omega _g=-{\frac{{8\pi ^2}}{{45}}}T^4V+{\frac{{4}}9}T^2C.
\end{equation}

The total $\Omega $ can be found from summing the terms above, and other
thermodynamical quantities like the free energy and the internal energy can
be derived. For 3 massless quark flavors of equal chemical potential one
finds 
\begin{eqnarray}
\Omega =\left( -{\frac{{19\pi ^2}}{{36}}}T^4-{\frac 32}\mu ^2T^2-{\frac{{3}}{%
{4\pi ^2}}}\mu ^4+B\right) V+\left( {\frac{{41}}{{72}}}T^2+{\frac{{3}}{{8\pi
^2}}}\mu ^2\right) C
\end{eqnarray}
\begin{eqnarray}
F=\left( -{\frac{{19\pi ^2}}{{36}}}T^4+{\frac 32}\mu ^2T^2+{\frac{{9}}{{4\pi
^2}}}\mu ^4+B\right) V+\left( {\frac{{41}}{{72}}}T^2-{\frac{{3}}{{8\pi ^2}}}%
\mu ^2\right) C
\end{eqnarray}
\begin{eqnarray}
E=\left( {\frac{{19\pi ^2}}{{12}}}T^4+{\frac 92}\mu ^2T^2+{\frac{{9}}{{4\pi
^2}}}\mu ^4+B\right) V-\left( {\frac{{41}}{{72}}}T^2+{\frac{{3}}{{8\pi ^2}}}%
\mu ^2\right) C.
\end{eqnarray}

Strangelets are in mechanical equilibrium at fixed temperature and baryon
number when $dF=0$, corresponding to 
\begin{eqnarray}
BV=\left( {\frac{{19\pi ^2}}{{36}}}T^4+{\frac 32}\mu ^2T^2+{\frac{{3}}{{4\pi
^2}}}\mu ^4\right) V-\left( {\frac{{41}}{{216}}}T^2+{\frac{{1}}{{8\pi ^2}}}%
\mu ^2\right) C
\end{eqnarray}
In this case one gets the following expressions for the thermodynamic
potential, free energy, internal energy and baryon number: 
\begin{equation}
\Omega =\left( {\frac{{41}}{{108}}}T^2+{\frac{{1}}{{4\pi ^2}}}\mu ^2\right) C
\end{equation}
\begin{equation}
F=\left( 3\mu ^2T^2+{\frac{{3}}{{\pi ^2}}}\mu ^4\right) V+\left( {\frac{{41}
}{{108}}}T^2-{\frac{{1}}{{2\pi ^2}}}\mu ^2\right) C
\end{equation}
\begin{equation}
E=4BV
\end{equation}
\begin{equation}
A=\left( \mu T^2+{\frac{{1}}{{\pi ^2}}}\mu ^3\right) V-{\frac \mu {{4\pi ^2}}
}C.
\end{equation}

Notice that the equations above can also be used in connection with bulk
SQM, for instance in an astrophysical context, by simply putting $C=0.$ An
external pressure can be accommodated by substituting $B+P_{{\rm external}}$
in place of $B.$

\begin{figure}[h!tb]
\centerline{\psfig{file=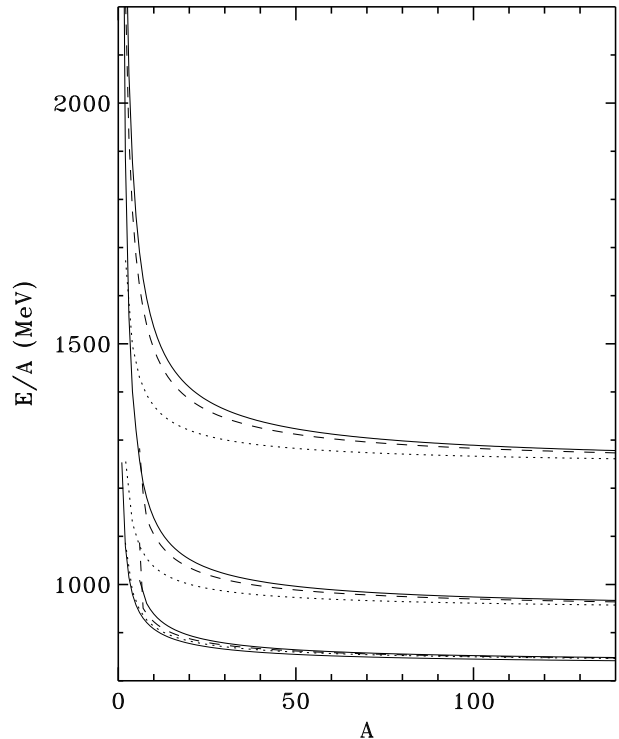,width=9cm}}
\caption{$E/A$ as a function of $A$ for strangelets with equal numbers
of massless $u$, $d$, and $s$ quarks for entropy per baryon of 0, 1, 5 and 10,
and $B^{1/4}=145$MeV. Solid curves include color singlet and zero
momentum constraints, dashed curves only the color singlet constraint,
and dotted curves are without constraints. Entropy increases upward. For
$S=0$ ($T=0$) the three curves completely overlap (lowest solid curve).}
\label{fig:ent}
\end{figure}

Dotted curves in Figure \ref{fig:ent} shows the energy per baryon for finite 
temperature strangelets
according to the formulae above. Results are given for fixed entropy
per baryon, where the entropy is calculated from $S\equiv -\partial \Omega
/\partial T|_{V,\mu }$. These results were first presented in
\cite{jenmad96}. A similar treatment, including finite $m_s$, was
published in \cite{heal96}, whereas Ref.\ \cite{musans96} shows results
for a corresponding finite temperature shell model calculation,
finding that shell structures
are washed away at $T>10\,$MeV, which means that liquid drop model and shell
model results become indistinguishable at high $T$ ($S/A$).

As discussed in more detail in 
\cite{jenmad96} further complications arise from the fact, that strangelets
must be color singlets. This has no influence on the ground state energy for 
$T=0$, but for $T>0$ quarks are statistically distributed over energy
levels, and the color singlet constraint reduces the number of possible
configurations, forcing the energy up for fixed entropy (see also
\cite{musans97}). The effect is
important for $A<100$ as illustrated in Figure \ref{fig:ent}. 
Similar effects result
from insisting that strangelets shall have a definite momentum. These
destabilising effects can be important in connection with experiments, which
inevitably create strangelets with rather high entropies. A tremendous
job remains to be done in calculating the details of strangelet
formation, evolution, and decay modes, including realistic 
non-equilibrium effects, etc.!

\section{SQM in cosmology}
\label{sec:cosmo}

\subsection{Formation, evaporation and boiling of quark nuggets}

If the cosmological quark-hadron phase transition was first order,
supercooling may result in concentration of baryon number inside shrinking
bubbles of quark phase. The amount of baryon concentration depends on the
permeability of the ``membrane'' separating the phases and on the turbulent
removal of quarks from the phase boundary. If a quark bubble is able to get
rid of entropy fast enough (primarily in the form of neutrinos and photons)
relative to the rate of baryon number removal, there is a chance of reaching
baryon number densities in the quark bubbles approaching nuclear matter
density. In other words, a quark nugget may form. Whether or not this
actually happens, or whether one is left with the less extreme, but also
interesting scenario where all of the quarks end up in inhomogeneously
distributed neutrons and protons, giving non-standard Big Bang
nucleosynthesis, has been a topic of much debate \cite
{wit84,apphog85,kur88,sumal90,jedful94b}, 
and the final word has probably not been said.

But even if cosmological quark nuggets do form, they find themselves in a
very hostile environment with a temperature of order 100 MeV. Under such
conditions the nuggets are unstable against surface evaporation \cite
{alcfar85,madhei86,sumkaj91,bhaal93} and boiling \cite
{alcoli89,madole91a,olemad93}; but the crucial question from a cosmological
point of view is whether some nuggets may survive due to the relatively
short time-scale for cooling the Universe (the age of the Universe at
temperature $T$ being roughly $t_{{\rm sec}}=T_{{\rm MeV}}^{-2}$).

Alcock and Farhi \cite{alcfar85} showed that the timescale for complete
evaporation of a quark nugget was smaller than the age of the Universe at
temperature $T$ for baryon numbers below 
\begin{equation}
A_{{\rm evap}}\approx 2\times 10^{56}\exp (-3I_n/T)f_n^3,
\end{equation}
where $f_n$ (the phase boundary penetrability of neutrons) was assumed to be
close to unity. For a homogeneous quark nugget the neutron binding energy $%
I_n=m_n-\mu _u-2\mu _d$ was estimated to be of order 20 MeV. For such a
binding, primordial nuggets with baryon number $A<10^{55}$ evaporate almost
instantly when neutrino heating becomes possible at $T\approx $ 50 MeV \cite
{alcfar85}.

However, the surface evaporation of neutrons and protons reduces $\mu _u$
and $\mu _d$, and leads to an increase in $\mu _s$. Weak decays, diffusion
and convection work to counteract this, but the net result is an $s$-quark
enriched layer near the surface. (Small nuggets are $s$-quark enriched
throughout their interior). The most efficient way to remove the $s$-quarks
is then to emit them in kaons ($\overline{K^0}$, $K^{-}$) along with thermal 
$\bar{u}$ and $\bar{d}$. A quasi-equilibrium situation arises with an
effective $I_n\approx $ 350 MeV \cite{madhei86}. Thereby the baryon number
of nuggets surviving evaporation is reduced to $10^{46}$, and a proper
inclusion of reabsorption of emitted hadrons (a calculation that has so far
not been done) may reduce the number somewhat.

Cosmological nugget evaporation (time-reversed) is closely related to the
distillation mechanism proposed for strangelet production in relativistic
heavy-ion collisions \cite
{liusha84,grekoc87,greal88,shaal89,gresto91,baral91,crades92,spial96}. There
strangeness enhancement occurs due to emission of $K^{+}$ and $K^0$.

The calculations described above assume that the penetrability of the phase
boundary is near 100\%. It has been argued that the penetrability may be
reduced by a few orders of magnitude in a chromoelectric flux tube model.
This would decrease $A_{{\rm evap}}$ by a factor $f_n^3$, permitting
small\-er nug\-gets (possibly down to $A=10^{39}$) to survive \cite
{sumkaj91,bhaal93}. Again, the limit on $A$ may be further reduced by
reabsorption.

Primordial nuggets are superheated, and may therefore boil by forming
bubbles of hadronic gas in their interiors \cite{alcoli89}. However, even
though boiling is thermodynamically allowed, it probably does not play an
important role for primordial nuggets (or in heavy-ion collisions for that
matter), since the time-scale is too short for bubble-nucleation to take
place \cite{madole91a,olemad93}. The surface evaporation described above is
thus the decisive mechanism.

Some authors have argued \cite{leelee91,cholee94}, 
that boiling will take place unless a large external
pressure (e.g.\ due to a gravitationally bound shell of nucleons) is there
to prevent it. Such gravitational stabilization only works for masses close
to those of stars ($A\approx 10^{57}$). However, the authors discuss only
whether boiling is thermodynamically possible, but neglect that there is not
enough time for the bubbles to nucleate.

Apart from trace abundances, one should not expect nuggets smaller than $%
10^{30}-10^{40}$ to survive from the early Universe. This however brings one
well within the causality limit set by the baryon number inside the horizon
during the cosmic quark-hadron phase transition, 
\begin{equation}
A_{{\rm hor}}\approx 10^{49}\left( \frac{{100{\rm MeV}}}T\right) ^2,
\end{equation}
and includes the ``most probable'' range of baryon numbers originally
predicted by Witten \cite{wit84}. It also leaves open the possibility that
SQM may explain the dark matter problem, and if we understood the details of
the quark-hadron phase transition, we could even calculate the relative
abundances of dark and ordinary matter from first principles.

There is a possibility, that also small traces of primordial nuggets with
low baryon numbers are left over from the early Universe. Even such traces
may in fact be ``observed'' using the astrophysical detectors discussed in
Section \ref{sec:cosmic}, or via Big Bang nucleosynthesis, 
as explained in Section \ref{subsec:bbn}.

\subsection{Quark nuggets and Big Bang nucleosynthesis}
\label{subsec:bbn}

A crucial property of quark nuggets is the positive electrostatic surface
potential of the quark phase, which is due to the quarks being stronger
bound than the electrons (electrostatic forces are weaker than strong
forces). For typical nugget parameters the electrostatic potential can
be several MeV, so except at very high temperatures, protons and nuclei are
repelled from nuggets, whereas neutrons are absorbed, adding one unit of
baryon number.

This opens the intriguing possibility of using SQM as an energy source \cite
{shaal89}, at least in principle. It also makes it possible to use Big Bang
nucleosynthesis as well as the properties of pulsars to place very stringent
limits on the abundance of quark nuggets in the Universe.

During Big Bang nucleosynthesis ($T\leq 1$ MeV), nuggets absorb neutrons but
not protons. This means that the presence of quark nuggets reduces the
neutron-to-proton ratio, thereby lowering the production of $^4$He. The
helium-production is very sensitive to the total amount of nugget-area
present, and in order not to ruin the concordance with observations, one
finds \cite{madrii85} that only nuggets with $A>A_{{\rm BBN}}\approx
10^{23}\Omega _{{\rm nug}}^3h^6f_n^3$ are allowed during nucleosynthesis.
Here $\Omega _{{\rm nug}}$ is the present-day nugget contribution to the
cosmic density (in units of the critical density), $h$ is the Hubble
parameter in units of 100 km sec$^{-1}$ Mpc$^{-1}$, and $f_n\leq 1$ is the
penetrability of the nugget surface. Slightly stronger limits can be
obtained from inclusion of all light nuclei instead of $^4$He only \cite
{ole90}. (Ref.\ \cite{schdel85} found good
correspondence with nucleosynthesis for a nugget-dominated, $\Omega =1$
Universe if $A\approx 10^{17}$, but as shown in
\cite{madhei86}, this was due to an erroneous emission rate for nucleons.)

The nucleosynthesis calculations leading to $A_{{\rm BBN}}$ neglected
inhomoge\-nei\-ties in the nucleon distribution, and all nuggets were
assumed to have the same baryon number. However, the formation of $^4$He is
an on-off process over a limited range of $A$, so the detailed behavior of
the inhomoge\-nei\-ties may not be so important.

Note that SQM, in spite of it carrying baryon number, does {\it not\/}
contribute to the usual nucleosynthesis limit on $\Omega_{{\rm baryon}}$.
The SQM baryon number is ``hidden'' in quark nuggets long before Big Bang
nucleosynthesis begins, and the nuggets only influence nucleosynthesis if
they have a big total surface area, as described above.

Evaporating nuggets would lead to strongly inhomogeneous nucleosynthesis
with enhanced heavy-element formation. This aspect has recently been studied
in \cite{jedful94a}.

\subsection{Quark nuggets as dark matter}

Witten \cite{wit84} argued that quark nuggets might be a natural explanation
of the cosmological dark matter problem, in principle allowing a calculation
of the relative amount of dark matter and ordinary baryons. In view of the
evaporation discussed above, this idea now seems less likely, but is
certainly not ruled out for $A>10^{30}$. Massive quark nuggets decouple from
thermal equilibrium with the radiation bath very early in the history of the
Universe, quickly slow down, and behave as cold dark matter in the context
of galaxy formation.

Of course it should again be noted, that all of the interesting cosmological
consequences of the quark-hadron phase transition require the transition to
be first order, in agreement with recent lattice QCD calculations.

\section{SQM in neutron stars; strange stars}
\label{sec:star}

It has been known for many years, that neutron stars may in fact be ``hybrid
stars'' consisting of ``ordinary'' nuclear matter in the outer parts and
quark matter in the central regions. This will be the case if SQM is
metastable at zero pressure, being stabilized relative to hadronic matter by
the high pressure within a neutron star \cite{baychi76,chanau76,fremcl78}.

If SQM is absolutely stable at zero pressure, an even more intriguing
possibility opens up, namely the existence of ``strange stars'' \cite
{wit84,haezdu86,alcfar86a,alc91} 
consisting completely of SQM (perhaps apart from
a minor crust to be discussed below). Such strange stars behave quite
differently from neutron stars due to the unusual
equation of state. For massless quarks the total energy density is given by $%
\rho =\rho _q+B$, and the total pressure by $P=P_q-B$, where $\rho _q$ is
the energy density of quarks, and the pressure of the quarks is $P_q=\rho
_q/3$, since massless quarks are relativistic. The equation of state is thus
given by 
\begin{equation}
P={\frac 13}\left( \rho -4B\right) .
\end{equation}
The exact equation of state taking into account $m_s\neq 0$ is very similar 
\cite{alcfar86a} since $s$-quarks are relativistic for low $m_s$ and not
present for high $m_s$. (I here assume that $\alpha _s=0$,
but recall from Section \ref{sub:bulksqm}, that a non-zero $\alpha _s$
effectively corresponds to a reduction of $B$).

The structure of a strange star is calculated from the Oppenheimer-Volkoff
equation, describing the balance between gravity and pressure gradient,
using the equation of state given above. The surface of the star corresponds
to $P=0$, a condition fulfilled for $\rho=4B$, which for typical values of $%
B $ is somewhat more than the density of ordinary nuclear matter! For
stellar masses below $1M_\odot$ ($M_\odot$ is the solar mass) this density
is almost constant throughout the star, so to a good approximation total
mass and radius are related by $M\propto R^3$, a relation in striking
contrast to ordinary neutron stars, where $M\propto R^{-3}$. This means that
low-mass neutron stars and strange stars have widely different radii,
possibly allowing observational distinction. Unfortunately Nature prefers to
form these compact objects with masses near $1.4 M_\odot$, according to
stellar evolution models. For such a mass gravity rather than bag pressure
plays the dominant stabilizing role, and there is no significant difference
between neutron star and strange star radii. Also the maximum mass given by
gravitational instability (the Chandrasekhar limit) is similar, of order $%
2M_\odot$. In contrast to ordinary neutron stars, which are unstable for
masses below $0.1\,M_\odot$, strange stars have no minimum mass; the
sequence continues smoothly to the domain of strangelets.

For the simple equation of state discussed above, the only natural energy
scale in the problem is $B^{1/4}$. Thus there exists a homology
transformation between strange star models for different values of $B$. In
particular, the maximum mass of a strange star is given by 
\begin{equation}
M_{{\rm max}}=2.006B_{145}^{-1/2}M_{\odot }.
\end{equation}
The corresponding minimal radius, maximal moment of inertia, maximal central
density, surface density, and minimal rotation period (the so-called Kepler
period corresponding to mass-shedding at the equator), are given by 
\begin{equation}
R_{{\rm min}}=10.94B_{145}^{-1/2}{\rm km},
\end{equation}
\begin{equation}
I_{{\rm max}}=2.256\times 10^{45}B_{145}^{-3/2}{\rm g~cm}^2,
\end{equation}
\begin{equation}
\rho _{{\rm max}}=1.97\times 10^{15}B_{145}{\rm g~cm}^{-3},
\end{equation}
\begin{equation}
\rho _{{\rm surf}}=4.102\times 10^{14}B_{145}{\rm g~cm}^{-3},
\label{rhosurf}
\end{equation}
\begin{equation}
P_{{\rm min}}=0.66B_{145}^{-1/2}{\rm ms}.  \label{pmin}
\end{equation}

Bare strange stars (strange stars with quark matter all the way to the
surface) have quite unusual properties. The density abruptly jumps from 0 to 
$\rho _{{\rm surf}}$ (Eq.~(\ref{rhosurf})), and the density is almost
constant through the interior (except when the mass is close to $M_{{\rm max}%
}$). The plasma frequency of the star is huge, meaning that photons with
energies below 20MeV are reflected from the surface, whereas the star itself
can only emit photons with higher energies \cite{alcfar86a,chmhae91}. Even
more important, because of the strong interaction binding of the surface
material, the star is not subject to the ``Eddington limit'', which for
ordinary neutron stars limits the luminosity to be below $10^{38}$ erg/s
(for higher luminosities the radiation pressure would exceed the
gravitational attraction and expel the surface layers). As discussed below,
this could lead to important ``applications'' of strange stars.

This approach may however be oversimplified because real strange stars may
have surfaces more like ordinary neutron stars. In particular, a solid crust
of ordinary material may form from accretion by the strange star after
formation, or from material that was not converted during neutron star
burning (see Section \ref{subsec:form}). Such a crust may be held up by the
extreme, outward directed electrostatic potential of $10^{17}$--$10^{18}$%
V/cm, created by the electron atmosphere with a thickness of a few hundred
Fermi. This atmosphere merely expresses that the electrostatic binding of
electrons is weaker than the strong binding of quarks; therefore the
electron distribution does not end abruptly like that of quarks (the
detailed structure was found from a Thomas-Fermi calculation by Alcock,
Farhi and Olinto \cite{alcfar86a}; see also \cite{ketweb95}).

The electrostatic potential can sustain a significant crust of ordinary
neutron star material. The limit is given by the neutron drip density ($%
4\times 10^{11}$ gcm$^{-3}$), above which neutrons drip out of nuclei and
would be swallowed by the quark phase. This crust may be decisive for
interpretation of pulsar behavior (Section \ref{subsec:puls}).

As emphasized by Glendenning, Kettner and Weber \cite{gleket95a,gleket95b}, 
the existence of
crusts not only changes the mass-radius relation for strange stars, but also
opens a rich plethora of new stellar configurations. In particular, one may
have a sequence of ``strange dwarfs'', much like white dwarfs except for an
SQM core. At present there is no well-studied model for formation of such
strange dwarfs.

Another possibility for formation of a (solid?) crust has been suggested
in \cite{benhor90}. This mechanism relies on
the existence of stable, low-baryon number strangelets (in this context
sometimes denoted ``quark-alphas'' for the $A=6$ strangelet analog of a
helium nucleus \cite{mic88}) which could act as ``nuclei'' in the surface
region. Whereas this possibility may seem less likely from the discussion in
Section \ref{sec:strangelet}, it can not be entirely ruled out.

Finally, it is worth noticing that Glendenning \cite{gle92} has argued that
neutron stars may contain regions with mixed quark and hadron phases.
(This possibility was missed in earlier studies due to an erroneous
assumption of {\em local} rather than {\em global} charge neutrality).
Depending on parameters the mixed phase region can occupy a significant
fraction of the star, and may show unusual topologies (plate-like or
cylinder-like structures, rather than just spherical quark bubbles embedded
in hadrons or vice versa \cite{heipet93a,glepei95,chrgle97}).

Studies of strange stars have not been pursued to the degree of detail known
for ordinary neutron stars, and it is premature to draw any detailed
conclusions. However, in the following, I shall look at some of the properties
expected and emphasize the possible observable differences between neutron
stars and strange stars.

\subsection{Neutrino cooling}

A distinction between strange stars and neutron stars was for a long time
believed to be a much more rapid cooling of SQM due to neutrino emitting
weak interactions involving the quarks \cite{alcfar86a}. Thus a strange star
was presumed to be much colder than a neutron star of similar age, a
signature potentially observable from x-ray satellites. Only a few
speculative mechanisms, such as the existence of kaon condensates might
mimic the speed of quark matter neutrino cooling. Recently the story has
been complicated considerably by the finding that ordinary neutron $\beta $%
-decay may be energetically allowed in nuclear matter \cite{latal91}, so
that the cooling rate can be comparable to that of SQM. For this reason I
shall not discuss the issue here, but refer the reader to an excellent
review of neutron star cooling by Pethick \cite{pet92}, and a recent
reinvestigation of strange star cooling by Schaab {\it et al.}\
\cite{schal97}.

\subsection{Pulsar glitches}
\label{subsec:puls}

One important feature seems to distinguish strange stars from neutron stars
in a manner with observable consequences, and that is the distribution of
the moment of inertia inside the star. Ordinary neutron stars older than a
few months have a crust made of a crystal lattice or an ordered
inhomogeneous medium reaching from the surface down to regions with density $%
2 \times 10^{14}\,{\rm g\,cm}^{-3}$. This crust contains about 1\% of the
total moment of inertia. Strange stars in contrast can only support a crust
with density below the neutron drip density ($4.3 \times 10^{11}\,{\rm g\,cm}%
^{-3}$). This is because free neutrons would be absorbed and converted by
the strange matter. Such a strange star crust contains at most a few times $%
10^{-5}$ of the total moment of inertia. This is an upper bound, since the
strange star may have no crust at all, depending on its prior evolution.
And recent studies of the mechanical balance between electric and
gravitational forces on the crust indicate, that only densities up to
perhaps $10^{11} {\rm g\,cm}^{-3}$ may be achieved \cite{hualu97,lu98}.

As stressed by Alpar \cite{alp87}, and also pointed out by Haensel, Zdunik,
and Schaeffer \cite{haezdu86}, and by Alcock, Farhi, and Olinto \cite
{alcfar86a}, this difference in the moment of inertia stored in the crust of
neutron stars and strange stars seems to pose significant difficulties for
explaining the glitch-phenomenon observed in radio pulsars with models based
on strange stars. Glitches are observed as a sudden speed-up in the rotation
rate of pulsars. The fractional change in rotation rate $\Omega $ is $\Delta
\Omega /\Omega \,\approx \,10^{-6}$---$10^{-9}$, and the corresponding
fractional change in the spin-down rate $\dot{\Omega}$ is of order $\Delta 
\dot{\Omega}/\dot{\Omega}\,\approx \,10^{-2}$---$10^{-3}$. Regardless of the
detailed model for the glitch phenomenon these jumps must involve the
decoupling and recoupling of a component in the star containing a
significant fraction, $I_i/I$, of the total moment of inertia; $%
fI_i/I\,=\,\Delta I/I\,\approx \,\Delta \Omega /\Omega \,\approx \,10^{-6}$%
---$10^{-9}$ (Alpar actually argued that $fI_i/I\,\approx \,\Delta \dot{%
\Omega}/\dot{\Omega}\,\approx \,10^{-2}$---$10^{-3}$, where $f$ is the
fractional change in $I_i$, but this is not necessary \cite{gleweb92}). This
role is played by the inner crust of an ordinary neutron star, but the crust
around a strange star is smaller; less than a few times $10^{-5}M_{\odot }$
with $I_{{\rm crust}}/I$ around a few times $10^{-5}$ for ordinary neutron
star masses of $1.4M_{\odot }$ (higher for less massive stars). These
numbers are based on models by Glendenning and Weber \cite{gleweb92}
assuming a maximum mass crust, i.e.\ a crust reaching neutron drip density
at the base, so it seems fair to conclude, that strange stars in fact may
have sufficiently massive crusts to account for glitches, but that
parameters in that case are fairly tightly constrained.

Other possibilities for glitches in strange stars could involve a crust
composed of strangelets (cf.\ the ``quark-alpha'' scenario in
\cite{benhor90}), not to mention the
possibility of a quark-hadron mixed phase 
\cite{gle92,heipet93a,glepei95,chrgle97}.
There is still a lack of any detailed model for how the magnetic field
structure and other crucial aspects of a pulsar can be modeled for strange
stars. Presumably a strange star cannot do the job without significant
structure, such as a crust and/or superfluidity/superconductivity in certain
regions. These issues have only been very superficially studied and need
further consideration. The present lack of
such models should not be used to dismiss the possibility of strange stars.

\subsection{Strange star oscillation and maximum rotation rate}

One of the most interesting differences between neutron stars and strange
stars is related to the damping of instabilities. 

First it should be noticed
that a strange star is a very stable system. Strange stars may have radial
oscillations with a fundamental period of 0.06--0.3\thinspace ms \cite
{datal92}, but these are characterized by rapid damping in a matter of
seconds \cite{wanlu84,saw89,cutlin90,mad92}. This is due to the extremely
high viscosity of SQM.

The large viscosity also plays a role in setting the maximum rotation limit
for strange pulsars (or hybrid stars with SQM cores). The ultimate rotation
limit corresponds to mass-shedding from the equator of the star (this is
called the Kepler limit and is of order 0.6 msec for a strange star,
Eq.\ (\ref{pmin}); see Zdunik \cite{zdu91} for a review). But before reaching
such rotation rates, the pulsars become unstable to non-radial deformations
and are slowed down by emission of gravitational radiation. Shear and bulk
viscosities tend to stabilize the star against these instabilities \cite
{saw89,colmil92}, and the high value for the bulk viscosity may mean that
strange pulsars in contrast to ordinary pulsars can reach submillisecond
periods \cite{mad92}. Thus the discovery of very fast pulsars may be an
indication favoring the existence of strange stars.

And even more exciting, it has been shown over the last few months, that
ordinary neutron stars when they are young and hot are subject to a new
class of instabilities, called r-mode instabilities 
\cite{and98,frimor98,linowe98}, which
during their first year of existence slows the rotation rate to
only a few per cent of the Kepler limit. Rotation periods faster than 10
msec are unlikely after that, until some pulsars at a much later age may
be spun-up by angular momentum transfer in binary systems, and thereby
explain the rapid old pulsars with periods down to 1.56 msec. 
In contrast, strange stars are not subject to these instabilities until
they are thousands of years old, and even then only for periods faster
than 2--3 msec \cite{mad98}. This seems to imply, that the most robust
signature for the existence of strange stars (or neutron stars with
a substantial fraction of high viscosity quark matter in the interior)
is to search for young pulsars with rotation periods below, say, 5 msec
(even stars with longer periods may candidate). These can not be
ordinary neutron stars, whereas quark matter is the only substance known
to have a bulk viscosity high enough to offer an explanation.

The bulk viscosity of strange quark matter depends on the rate of the
non-leptonic interaction 
\begin{equation}
u+d\leftrightarrow s+u.  \label{conver}
\end{equation}
(The rate for this reaction has recently been calculated by Madsen \cite
{mad93c}, and Heiselberg \cite{hei92}; earlier studies, including that of
Ref.\ \cite{heimad86} are incorrect). This
reaction changes the concentrations of down and strange quarks in response
to the density changes involved in vibration or rotational instabilities,
thereby causing dissipation. This dissipation is most efficient if the rate
of reaction (\ref{conver}) is comparable to the frequency of the density
change. If the weak rate is very small, the quark concentrations keep their
original values in spite of a periodic density fluctuation, whereas a very
high weak rate means that the matter immediately adjusts to follow the true
equilibrium values reversibly. But in the intermediate range dissipation due
to $PdV$-work is important.

The importance of dissipation due to Eq.~(\ref{conver}) was first stressed
by Wang and Lu \cite{wanlu84} in the case of neutron stars with quark cores.
These authors made a numerical study of the evolution of the vibrational
energy of a neutron star with an $0.2M_{\odot }$ quark core, governed by the
energy dissipation due to Eq.~(\ref{conver}). Sawyer \cite{saw89} expressed
the damping in terms of the bulk viscosity, a function of temperature and
oscillation frequency, which he tabulated for a range of densities and
strange quark masses. Sawyer's tabulation has later been used in studies of
quark star vibration \cite{cutlin90}, and of the gravitational radiation
reaction instability determining the maximum rotation rate of pulsars \cite
{colmil92}. The latter study concluded, that the bulk viscosity is large
enough to be important for temperatures exceeding 0.01 MeV, but that it
should be a few orders of magnitude larger to generally dominate the
stability properties.

However, as has been pointed out in \cite{mad92}, the bulk
viscosities calculated in \cite{saw89} depend on the assumption, 
that the rate of
Eq.~(\ref{conver}) can be expanded to first order in $\delta \mu =\mu _s-\mu
_d$, where $\mu _i\approx 300$MeV are the quark chemical potentials. This
assumption is not correct at low temperatures ($2\pi T\ll \delta \mu $),
where the dominating term in the rate is proportional to $\delta \mu ^3$.
Furthermore, the rate in \cite{saw89} is too small 
by an overall factor of 3, and a
discrepancy of 2--3 orders of magnitude, perhaps due to unit conversions,
appears as well. Taken together, these effects lead to an upward correction
of the bulk viscosity by several orders of magnitude, and thereby increases
the importance for the astrophysical applications. The non-linearity of the
rate also means, that the bulk viscosity is no longer independent of the
amplitude of the density variations. The resulting bulk viscosity is (in
cgs-units, with $m_s$, $T$, and $\mu _d\approx 235{\rm MeV}(\rho /\rho _{%
{\rm nuc}})^{1/3}$ in MeV, and the oscillation frequency $\omega $ in s$^{-1}
$) 
\begin{eqnarray}
\zeta \approx 3.09\times 10^{28}m_s^4\omega ^{-2}\left( {\frac \rho {\rho _{%
{\rm nuc}}}}\right) \left[ {\frac 34}\left( {\frac{m_s^2}{{3\mu _d}}}{\frac{{%
\Delta v}}{{v_0}}}\right) ^2+4\pi ^2T^2\right] {\rm g\,cm}^{-1}{\rm s}^{-1}.
\label{bulkvis}
\end{eqnarray}
For typical values ($m_s=100$ MeV, $\mu _d=300$MeV, $\omega =2\times 10^4$ s$%
^{-1}$) this is $\zeta \approx 1.6\times 10^{28}\left[ 93(\Delta
v/v_0)^2+39T^2\right] {\rm g\,cm}^{-1}{\rm s}^{-1}$, where $\Delta v/v_0$ is
the perturbation amplitude.

For a star of constant density (an excellent approximation for a strange
star, except very close to the gravitational instability limit) Sawyer \cite
{saw89} estimated the damping time as 
\begin{equation}
\tau _D\approx 1.5\times 10^{25}\zeta ^{-1}{\rm s}.
\end{equation}
Thus, even at very low temperatures, high amplitude oscillations are damped
in fractions of a second, and those of low amplitude in a matter of minutes,
if one takes into account, that the temperature of the star increases due to
the heat released by viscous dissipation, which can speed up the damping of
vibrations.

The discussion above was based on rather crude estimates \cite{mad92}. A
detailed, general relativistic, numerical treatment along the lines of
Cutler {\it et al.} \cite{cutlin90} is clearly needed.

As mentioned previously, viscosity also plays an important role 
in setting the maximum rotation rate
of pulsars. Gravitational radiation reaction instabilities (as opposed to
``Keplerian mass-shedding'') is supposed to set the ultimate 
rotation rate limit, but
the larger the damping by shear and bulk viscosity is, the closer the rate
can get to the Keplerian limit given in Eq.\ (\ref{pmin}).

The shear viscosity of SQM due to quark scattering has recently been
recalculated by Heiselberg and Pethick \cite{heipet93b}. Their results for 
$T\ll \mu $ can be written as 
\begin{equation}
\eta \approx 4.0\times 10^{15}\left( {\frac{{0.1}}{{\alpha _S}}}\right)
^{5/3}\left( {\frac \rho {{\rho _{nuc}}}}\right) ^{14/9}T^{-5/3}{\rm gcm}%
^{-1}{\rm s}^{-1}.
\end{equation}
Investigations by Colpi and Miller \cite{colmil92} based on the older
viscosities in \cite{saw89,haejer89}
indicated, that the minimal rotation period of strange stars might be set by
the gravitational radiation reaction instability of $m=2$ or $m=3$ modes at
or just below 1 millisecond. With the new, much larger, viscosities, the
non-axisym\-metric instabilities will be suppressed, and it is not
unreasonable to expect, that the maximum rotation frequency of strange stars
will be close to the Keplerian limit. Detailed numerical calculations like
those in Colpi and Miller \cite{colmil92}, including the new viscosities and
effects of dissipative heating, are required to settle the issue, but they
are complicated by the non-linear behavior of the new bulk viscosity.

Whether or not the ultimate rotation period of strange stars can be
significantly smaller than for neutron stars is of importance for old
pulsars spun-up by accretion. But perhaps the most clear-cut signature
for the existence of strange stars would be the (almost) lack of
sensitivity to r-mode instabilities, which as mentioned earlier allows
young strange stars to rotate much faster than young neutron stars
\cite{mad98}.

\subsection{Gamma-ray bursters}

Strange stars because of their high surface density, strong binding (making
it possible to circumvent the Eddington limit), and special emission
properties have been suggested as explanations for some of the more
mysterious cosmic events, namely $\gamma$-ray bursters. These are bursts of $%
\gamma$-rays of a few seconds duration, coming from unidentified sources
which are presumably at extragalactic distances.

No consensus exists concerning the nature of these bursts, but Alcock, Farhi
and Olinto \cite{alcfar86b} suggested a detailed model for the most
prominent of the bursters, the one on 5 March 1979. Their model is based on
an impact of a $10^{-8}M_{\odot }$ lump of SQM on a rotating strange star,
and the authors are able to explain most of the observations concerning
energetics and time-scales under the assumption that the burster is located
in a supernova remnant in the Large Magellanic Cloud, as position
measurements seem to indicate. An alternative model for this source and for
soft $\gamma $-repeaters in the framework of strange stars with
``quark-alpha'' surface properties was suggested in \cite{horvuc93}.
Other strange star models for soft $\gamma $-repeaters and $x$-ray
bursters include \cite{chedai98,cheal98}.

$\gamma $-ray bursters at truly cosmological distances could be due to
collisions of two strange stars in binary systems \cite{haepac91}, each
collision releasing $10^{50}$ ergs in the form of gamma rays over a
time-scale of 0.2 s.

There are, however, literally hundreds of different models for
$\gamma$-bursts, and in spite of improved observational data the
interpretation is at present unclear.

A recent identification of the $x$-ray source Her X-1 as a strange
star \cite{lidai95} was unfortunately based on incorrect use of bag
model parameters \cite{mad96}.

\subsection{Formation of strange stars}
\label{subsec:form}

If strange quark matter is stable, strange stars may be formed during
supernova-explosions, and neutron stars can be converted to strange stars by
a number of different mechanisms, such as pressure-induced transformation to
uds-quark matter via ud-quark matter, sparking by high-energy neutrinos, or
triggering due to the intrusion of a quark nugget. These and other
possibilities were described by Alcock, Farhi, and Olinto \cite{alcfar86a}.

As soon as a lump of strange matter comes in contact with free neutrons it
starts converting them into strange matter. The burning of a neutron star
into a strange star was discussed by Baym {\it et al.} \cite{bayal85} and
Olinto \cite{oli87}, and it was shown that the star would be converted on a
rather small time-scale set by quark diffusion and flavor-changing weak
interactions. (The huge difference in the speed of the conversion front
found in these papers is partly due to the omission of a factor $c^{1/2}$,
where $c$ is the speed of light, in equation (6) of Baym {\it et al.}) Later
studies \cite{heibay91,olemad91} found burning times in the range of 1--10$^3$
seconds under various parameter assumptions (see also Olinto \cite{oli91}
for a review). For the fastest burning times, the energy liberated may be
important for the supernova mechanism and supernova neutrino bursts. Horvath
and Benvenuto \cite{horben88} have questioned the stability of ''slow''
neutron combustion and suggested that the conversion takes place much faster
as a detonation. So far, the investigations of neutron star burning have
been rather crude, neglecting many aspects of transport theory, heat
conduction etc. A detailed study of this phenomenon would be interesting.

Perhaps the most likely mechanisms for initiating the formation of a strange
star involves either a seed of SQM in the star (see Section \ref{sec:cosmic}),
or thermal formation of quark matter bubbles. Thermal triggering of
neutron star transformation may be understood qualitatively in terms of
simple boiling theory. Before considering a more realistic equation of state
it is instructive to study the boiling of a pure neutron gas into quarks.
The quark bubbles formed consist of $u$ and $d$ quarks in the ratio 1:2;
only later weak interactions may change the composition to an energetically
more favorable state. Thus quark chemical potentials are related by $%
\mu_d=2^{1/3}\mu_u$, and $\mu_n=\mu_u+2\mu_d=(1+2^{4/3})\mu_u$, assuming
chemical equilibrium across the phase boundary.

The free energy involved in formation of a spherical quark bubble of radius $%
R$ and volume $V$ is given by 
\begin{equation}
F=-\Delta PV+8\pi \gamma R
\end{equation}
where 
\begin{equation}
\Delta P=P_{ud}-P_n={\frac{{\mu _u^4+\mu _d^4}}{{4\pi ^2}}}-B-P_n
\label{delP}
\end{equation}
and the curvature energy coefficient 
\begin{equation}
\gamma ={\frac{{\mu _u^2+\mu _d^2}}{{8\pi ^2}}}.
\end{equation}
The free energy has a maximum at the critical radius 
\begin{equation}
r_c=(2\gamma /\Delta P)^{1/2}
\end{equation}
and the corresponding free energy 
\begin{equation}
W_c\equiv F(r_c)=16\pi \gamma r_c/3
\end{equation}
is the work required to form a bubble of this radius which is the smallest
bubble capable of growing. It is a standard assumption in the theory of
bubble nucleation in first order phase transitions that bubbles form at this
particular radius at a rate given by\footnote{%
The prefactor may differ from $T^4$, but this is of minor practical
importance due to the dominant exponential.} 
\begin{equation}
{\cal R}\approx T^4\exp (-W_c/T).
\end{equation}

The simplest possible equation of state for the neutron gas is that of a
zero temperature, nonrelativistic degenerate Fermi-gas, where 
\begin{equation}
P_n={\frac{{(\mu_n^2-m_n^2)^{5/2}}}{{15\pi^2m_n}}}
\end{equation}
and the baryon density 
\begin{equation}
n_B={\frac{{(\mu_n^2-m_n^2)^{3/2}}}{{3\pi^2}}}
\end{equation}

\begin{figure}[h!tb]
\begin{tabular}{ll}
\hspace{-0.8cm}
\psfig{file=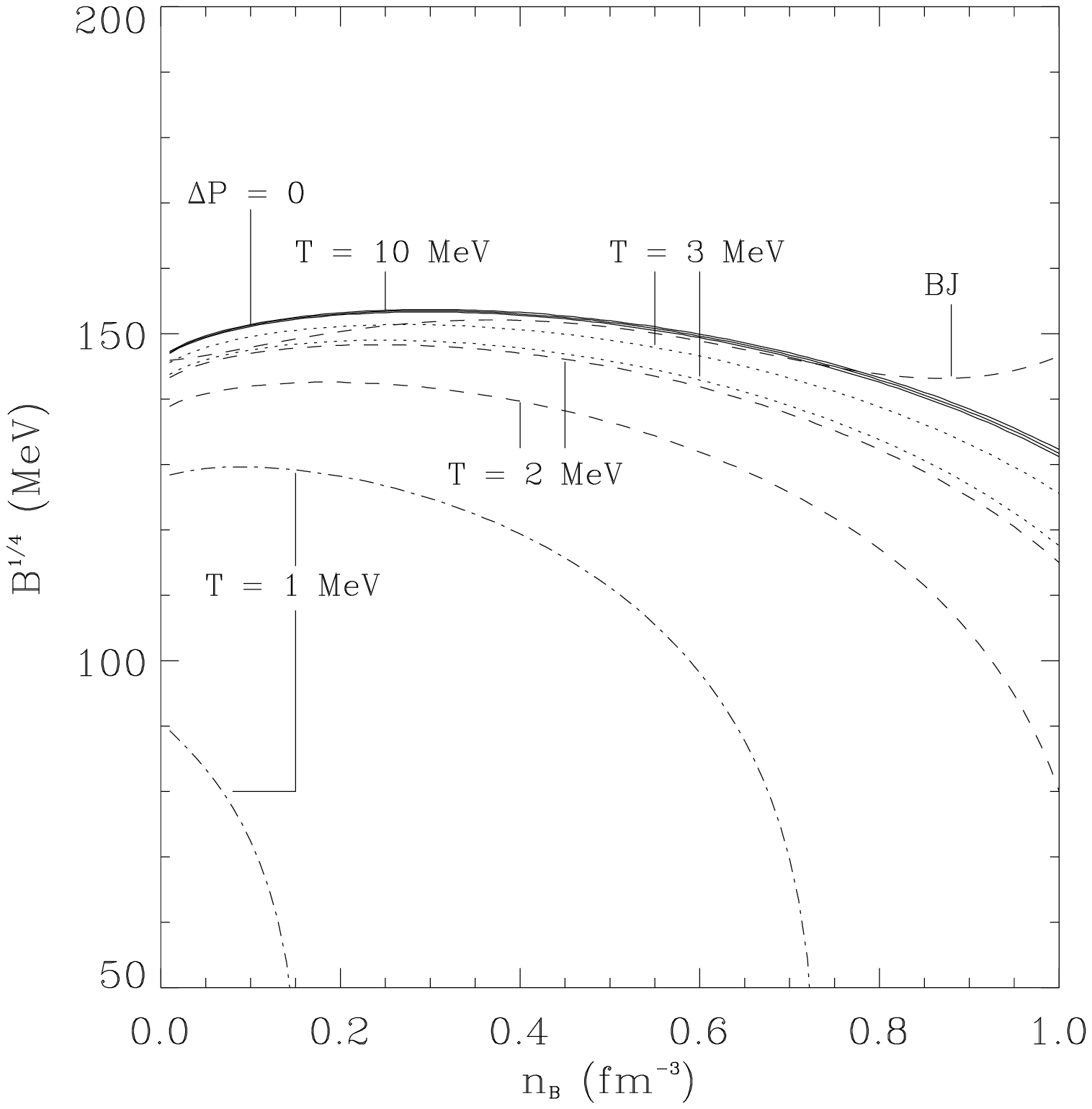,width=8.5cm}
\hspace{-1.3cm}
\psfig{file=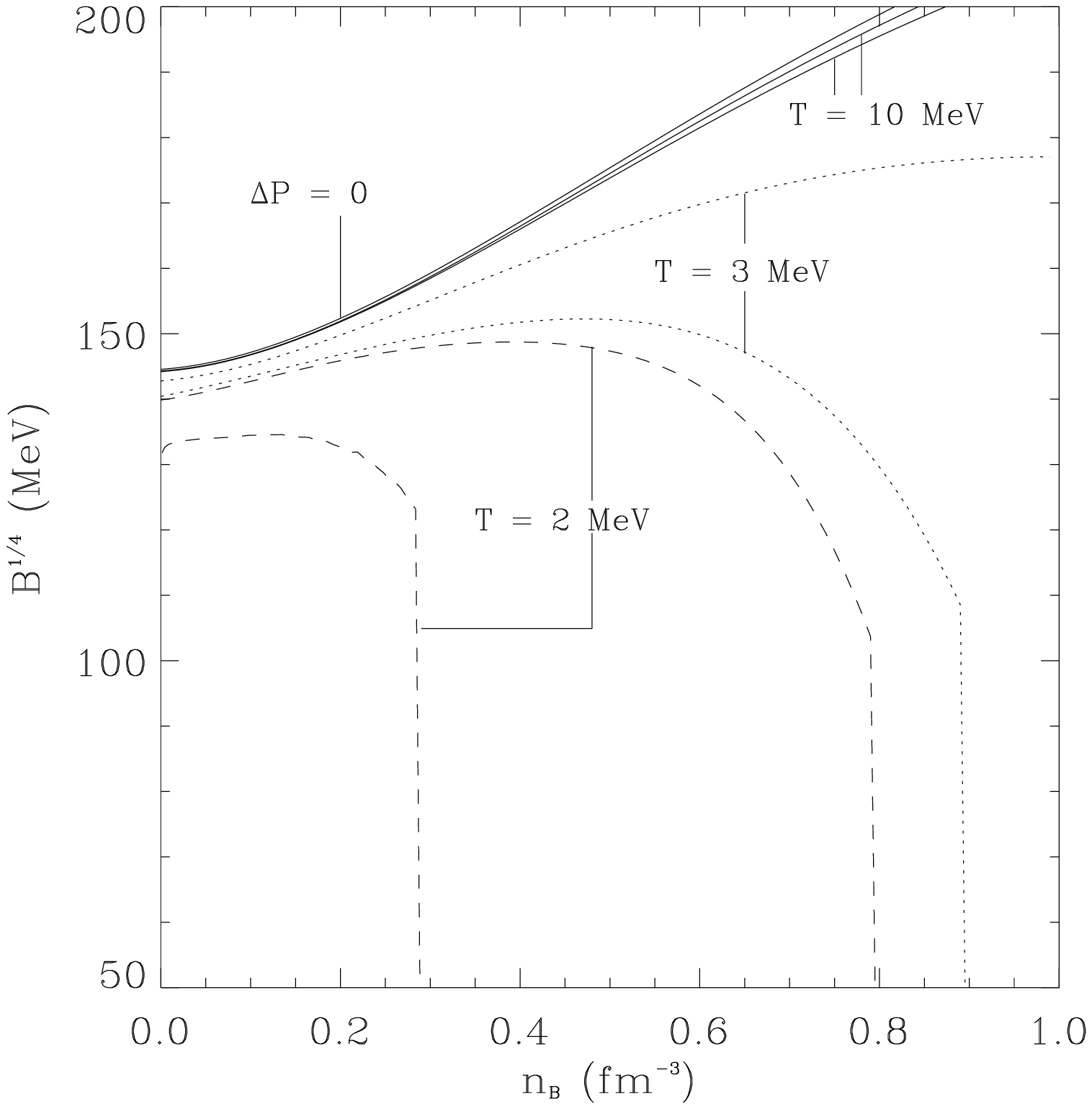,width=8.5cm}
\end{tabular}
\caption{Upper limits on the bag constant allowing thermal nucleation of
quark matter bubbles in neutron stars as a function of baryon number
density in the hadron phase. To the left are shown results for the
simple neutron gas model discussed in the text. To the right for a more
realistic mean field approximation. See text for further explanations.}
\label{fig:form}
\end{figure}

A necessary condition for boiling is that $\Delta P>0$. This leads to an
upper limit on the bag constant, $B_{{\rm max}}$, from Eq.~(\ref{delP}) as
illustrated in Fig.\ \ref{fig:form} (from \cite{olemad94}---the
corresponding limit for the Bethe-Johnson equation of state is shown for
comparison; it is seen to be very similar). This was also used as a
criterion for neutron star stability by Krivoruchenko and Martemyanov \cite
{krimar91}.

Also shown in Fig.\ \ref{fig:form} is the limit on the 
bag constant below which bubble
nucleation takes place at rates exceeding 1 km$^{-3}$Gyr$^{-1}$ and 1 cm$%
^{-3}$s$^{-1}$, respectively, for temperatures of 1, 2, 3 and 10 MeV ($B_{%
{\rm max}}$ can be considered as the limit for infinite temperature). One
notes that the possibility of bubble nucleation is fairly insensitive to the
temperature as soon as $T$ exceeds a few MeV, whereas thermally induced
bubble nucleation is impossible for $T<2{\rm MeV}$ (recall from Section 
\ref{sub:simple}
that the stability of ordinary nuclei against decay into quark matter
requires that $B>(145{\rm MeV})^4$). This confirms an estimate in 
\cite{horben92} (see also \cite{iidsat97,iid97}). 
The range of bag constants for which a hot
neutron star may transform into quark matter is thus roughly $145{\rm MeV}%
<B^{1/4}<155{\rm MeV}$.

Results for a more realistic mean field equation of state are also shown
in the Figure (see \cite{olemad94} for further details).
While the detailed numbers change, the overall
conclusion does not. Quark matter bubbles may nucleate (possibly followed by
burning of the star into SQM) in neutron stars/supernovae if the bag
constant is low, and if the temperature exceeds a few MeV (thus the process
is most likely during the supernova explosion itself). Should thermal
nucleation not take place, one of the other mechanisms mentioned above must
be relied on. Apart from seed-induced burning, all of these are likely to be
much less efficient than thermal nucleation.

\section{SQM in cosmic rays}
\label{sec:cosmic}

De R\'{u}jula and Glashow \cite{dergla84} argued that unusual meteor-events,
earth-quakes, etched tracks in old mica, in meteorites and in cosmic-ray
detectors might be used for observation of quark nuggets hitting the Earth
or its atmosphere. In particular they were interested in the possibility of
detecting a galactic dark matter halo of nuggets, where typical velocities
would be a few hundred kilometers per second, given by the depth of the
gravitational potential. Even if nuggets only survived from the Big Bang in
small numbers, or were spread in our galaxy by secondary processes such as
strange star collisions, there could be a potentially observable flux of
nuggets hitting the Earth. The only data actually investigated in their
paper came from a negative search for tracks in ancient mica, and
corresponded to a lower nugget flux limit of $8\times 10^{-19}\,{\rm cm}%
^{-2}\,{\rm s}^{-1}\,{\rm sr}^{-1}$, for nuggets with $A>1.4\times 10^{14}$
(smaller nuggets would be trapped in layers above the mica samples studied).
This can also be expressed as an excluded range of 
$1.4\times 10^{14}<A<8\times 10^{23}\rho _{24}v_{250},$
where $v\equiv 250{\rm km\,s}^{-1}v_{250}$ and $\rho \equiv 10^{-24}{\rm %
g\,cm}^{-3}\rho _{24}$ are the typical speeds and mass density of nuggets in
the galactic halo. The speed is given by the depth of the gravitational
potential of our galaxy, whereas $\rho _{24}\approx 1$ corresponds to the
density of dark matter. In these units the number of nuggets hitting the
Earth per ${\rm cm}^2$ per second per steradian is $6.0\times 10^5A^{-1}\rho
_{24}v_{250}$.

\begin{figure}[h!tb]
\centerline{\psfig{file=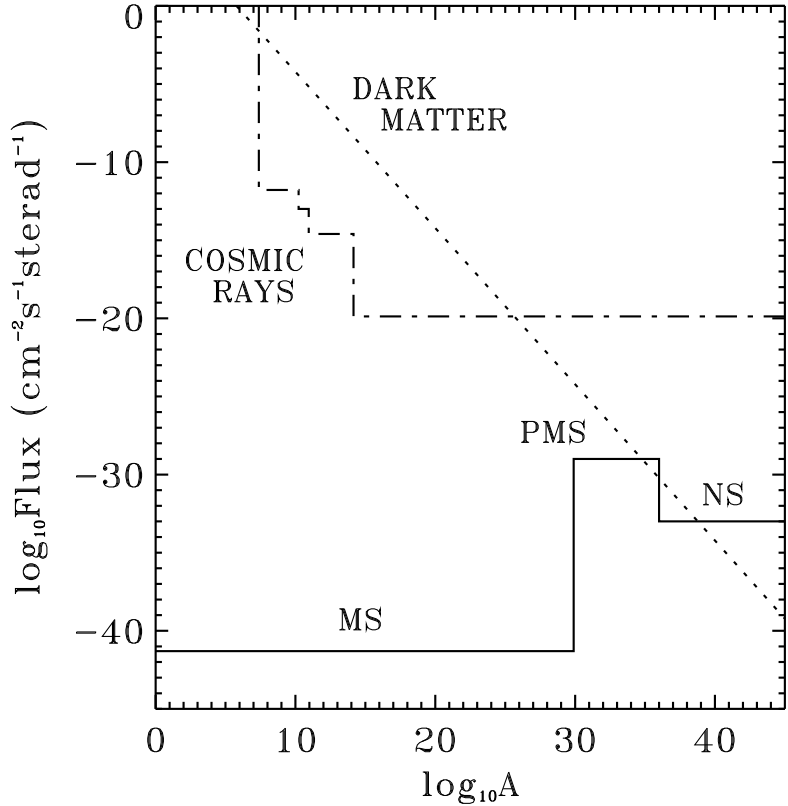,width=12cm}}
\caption{Astrophysical flux-limits \protect\cite{mad88} compared to the
flux expected for a galactic halo of nuggets being the dark matter 
\protect\cite{dergla84}, and to the experimental results for cosmic rays 
\protect\cite{pri88,low91}. 
The three horizontal parts of the solid curve correspond to
capture in main se\-quen\-ce su\-per\-no\-va progenitors, post main
se\-quen\-ce stars, and neutron stars younger than the Vela pulsar ($10^4$
years).}
\label{fig:cosmic}
\end{figure}

Later investigations have improved these flux limits somewhat.
These Earth-based flux-limits \cite{pri88,low91} are shown in 
Figure \ref{fig:cosmic}.
It is seen that quark nuggets with $3\times 10^7<A<5\times 10^{25}$ seem
incapable of explaining the dark halo around our galaxy, but a low flux
either left over from the Big Bang or arising from collision of strange
stars cannot be ruled out. If the strange matter hypothesis is valid, one
should indeed expect a significant background flux from stellar collisions,
since several pulsars are members of binary systems,
where the two components are ultimately going to collide. If such collisions
spread as little as 0.1$M_{\odot }$ of non-relativistic
strangelets with baryon number $A$, a
single collision will lead to a flux of $10^{-6}A^{-1}v_{250}$ cm$^{-2}$s$%
^{-1}$sterad$^{-1}$, assuming strangelets to be spread homogeneously in a
halo of radius 10 kpc.

Such a flux-level is below the sensitivity of present experiments, but
Madsen \cite{mad88} suggested that neutron stars and their stellar
``parents'' may be used as alternative large surface area, long integration
time detectors. The reason is simple. The presence of a single quark nugget
in the interior of a neutron star is sufficient to initiate a transformation
of the star into a strange star \cite{wit84,alcfar86a,bayal85}. The
time-scale for the transformation is short, between seconds and minutes \cite
{bayal85,oli87,oli91,heibay91,olemad91}, so observed pulsars would have been
converted long ago, if their stellar progenitors ever captured a quark
nugget, or if the neutron stars themselves absorbed one after formation.

The rate at which quark nuggets hit the surface of a star depends on the
phase space distribution of nuggets relative to the star. For an infinite
bath of positive energy nuggets with an isotropic, monoenergetic
distribution function, the number accretion rate is given by 
\begin{eqnarray}
F=1.39\times 10^{30\,}{\rm s}^{-1}\,A^{-1}{MR}\rho _{24}v_{250}^{-1}\left[
1\,+\,0.164v_{250}^2{RM}^{-1}\right] ,
\end{eqnarray}
where $M$ and $R$ denote the stellar mass and radius in solar units.
For the Sun the second term in parenthesis (the geometrical term)
contributes only slightly to the accretion rate, and the contribution is
even less important for more massive stars and for compact objects like
white dwarfs and neutron stars (in contrast, the geometrical term dominates
for accretion onto the Earth). In the following I therefore only take the
first term (gravitational) into account.

To convert a neutron star into strange matter a quark nugget should not only
hit a supernova progenitor but also be caught in the core. Similarly,
nuggets hitting a neutron star after its creation have to penetrate the
outer layers and reach the neutron drip region. These issues were discussed
in \cite{mad88}.

A main sequence star is capable of capturing quark nuggets with baryon
numbers below $A_{{\rm STOP}}$, where 
\begin{equation}
A_{{\rm STOP}}=5.0\times 10^{31}M^{-1.8}.
\end{equation}
This works for non-relativistic nuggets, which are basically braked by
inertia, i.\ e.\ they are slowed down by electrostatic scatterings after
plowing through a column of mass similar to their own, and afterwards settle
in the stellar core. In particular it is valid for nuggets moving with
virial speed in our galactic halo. Relativistic nuggets, like those reported
in some cosmic ray observations, may be destroyed after collisions with
nuclei in the stellar atmospheres, and so the limit can not be used
immediately, but it is worth noticing, that even a tiny fraction of a nugget
surviving such an event and settling in the star is sufficient to convert
the neutron star to a strange star.

For nuggets with $A<A_{{\rm STOP}}$ the sensitivity of main sequence stars
as detectors is remarkable, as it is given by the limit of one nugget
hitting the surface of the supernova progenitor in its main sequence
lifetime! Converted into a flux, ${\cal F}$, of nuggets hitting the Earth
per cm$^2$ per sec per steradian, it corresponds to 
\begin{equation}
{\cal F}=4\times 10^{-42}M^{0.1}v_{250}^2.
\end{equation}
As can be seen from Figure \ref{fig:cosmic}, 
this is a factor of $10^{20}$--$10^{40}$ more
sensitive than ordinary experiments!

{\it If} it is possible to prove that some neutron stars are indeed neutron
stars rather than strange stars, the sensitivity of the astrophysical
detectors rules out quark nuggets as being the dark matter for baryon
numbers in the range $A < 10^{34-38}$. And it questions the whole idea of
stable strange quark matter, since it seems impossible to avoid polluting
the interstellar medium with nuggets from strange star collisions or
supernova explosions at fluxes many orders of magnitude above the limit
measurable in this way.

{\it If} on the other hand SQM is stable, then all neutron stars are likely
to be strange stars, again because some pollution can not be avoided.

The Sun would in this way accrete $3.7\times 10^{-20}\rho
_{24}v_{250}^{-1}M_{\odot }/{\rm year}$, or a total of $10^{-10}\rho
_{24}v_{250}^{-1}M_{\odot }$ in its total lifetime on the main sequence.
Very low-mass nuggets collected near the solar center in this manner might
have an impact on the energy production \cite{jan88}, but the effect is
negligible unless the electrostatic barrier at the nugget surface is much
smaller than expected, or unless very special circumstances allow nuggets to
catalyze nuclear reactions \cite{takboy88}.

The Sun will develop into a white dwarf in about $6\times 10^9$ years.
As just mentioned, the Sun would accrete a core of $10^{-10}\rho
_{24}v_{250}^{-1}M_{\odot }$ in its total lifetime on the main sequence.
Such accretion is too small to lead to a
strange dwarf distinguishable from an ordinary white dwarf as suggested
in \cite{gleket95a,gleket95b}.
However, higher concentrations could occur if quark nuggets were
somehow mixed into the gas cloud from which the star originally formed.
Whether this is likely to happen depends strongly on assumptions regarding
the velocity distribution of the nuggets formed, and the possibility of
interactions with the gas \cite{calfri91}.

Most of the discussion above dealt with halo nuggets moving at
non-relati\-vis\-tic velocities. Relativistic nuggets are not as easily
detected using neutron stars, since they may be destroyed in collisions with
nuclei in the star. On the other hand two relativistic candidate events with
charge $Z=14$ and mass $A\approx 370$ were found in a balloon experiment by
Saito {\it et al.} \cite{saial90}. This corresponds to a rather high flux,
and it is not quite clear how to produce such nuggets, though spallation of
larger nuggets originating from strange star collisions may be involved \cite
{boysai93}. 

Quark nuggets have also been suggested as candidates for the Centauro
cosmic-ray events \cite{wit84,bjomcl79,halliu85}. Centauro primaries may
have a flux as high as $10^{-14}{\rm cm}^{-2}{\rm s}^{-1}$ and $A\approx
10^3 $. Since Centauro primaries move at relativistic speeds they are
destroyed by inelastic collisions when hitting a star, so the flux-limits
given above cannot directly be used to rule out quark nuggets as Centauro
primaries. However the mechanism producing the primaries must be tuned so
that it only produces relativistic quark nuggets in order not to conflict
with the flux-limits in Figure \ref{fig:cosmic} for non-relativistic nuggets.

\section{Conclusion and outlook}
\label{sec:concl}

The possible stability of strange quark matter is a fundamentally
exciting idea. Should it turn out to be true, many textbooks in nuclear,
particle and astrophysics will need revisions, but our daily lives will
not be changed dramatically, apart from possible technological
applications such as energy production and disposal of radioactive
waste \cite{shaal89,dessha91}.
There are two main reasons why stability of SQM is
possible without drastic consequences. The first reason is that
stability requires a certain minimum strangeness content, so ordinary
nuclei do not decay into strangelets. The second reason is the positive
electrostatic potential of the quark phase in a strangelet, which means
that you could walk around with a lump of SQM in your pocket without
being swallowed. 

While heavy-ion collisions is the way to look for small (meta)stable
strange\-lets, astrophysics gives a possibility for testing larger (and
therefore more stable) SQM-systems. Direct cosmic ray searches is an
obvious way to look for intermediate baryon numbers in the form of
relativistic or non-relativistic lumps produced in strange star
collisions, and for leftovers from the Big Bang. The latter
can only exist for very high baryon numbers (cf.\ Section \ref{sec:cosmo}), 
whereas a
galactic background of the former seems unavoidable if the strange
matter hypothesis is correct. 

Strange stars may be the most promising place to look for SQM, but as
explained in Section \ref{sec:star} it is 
actually hard to find clear-cut ways of
distinguishing strange stars from neutron stars, unless one finds an
object of very low mass. Pulsar rotation properties at present seem to
provide the best clue, in particular after the finding that young strange
stars in contrast to neutron stars are not braked by gravitational wave
emission due to r-mode instabilities.

If SQM is only metastable, heavy-ion physicists may still have a chance
of finding it; the cosmological quark-hadron phase transition may still
lead to inhomogeneities of importance for Big Bang nucleosynthesis
(without quark nuggets left over); and neutron stars may still have
strange matter cores.

In any case the confirmation or disproof of the existence of
(meta)stable strange quark matter via experiments and astrophysical
tests makes it possible to limit strong interaction parameters that are
otherwise difficult to probe. This in itself is a good reason for
continued studies of the physics and astrophysics of strange quark
matter.

\section*{Acknowledgments}
This work was supported in part by the Theoretical Astrophysics Center
under the Danish National Research Foundation.
I take this opportunity to thank the ``strangers'' among my
present and former PhD-students, Michael Olesen, Dan Jensen, Michael
Christiansen, and Gregers Neergaard for collaboration and many 
enlightening discussions.


\end{document}